\documentstyle[12pt]{report}

\newtheorem{lemma}{Lemma}
\def\blm{\begin{lemma}}
\def\elm{\end{lemma}}
\newtheorem{propos}[lemma]{Proposition}
\newtheorem{theorem}{Theorem}
\newtheorem {guess}{Conjecture}
\def\beq {\begin{equation}}
\def\eeq {\end{equation}}
\def\btab{\begin{tabbing}}
\def\etab{\end{tabbing}}
\def\bar{\begin{array}}
\def\ear{\end{array}}
\def\bea{\begin{eqnarray}}
\def\eea{\end{eqnarray}}
\def\cot{\otimes}
\def\bpp{\begin{propos}}
\def\epp{\end{propos}}
\def\btm{\begin{theorem}}
\def\etm{\end{theorem}}
\def\ben{\begin{enumerate}}
\def\een{\end{enumerate}}

\def \twoheadrightarrow {\to \mkern -15mu \to}

\def\id{{1\mkern-5mu {\rm I}}}

\def\noin{{\in \mkern-8mu | }}

\def\diff {{\cal D}i\!f\!f}

\def\isto {\widetilde{\longrightarrow}}
\def\bdot {\hbox{\raise .4ex\hbox{\large\bf .}}}
\def\ub{\underline}
\def\rcoa{\leftharpoondown}
\def\lcoa{\rightharpoonup}
\def\rac{\!\triangleleft\!}
\def\lac{\!\triangleright\!}

\def\A{{\cal A}}
\def\oA{\A ^o}
\def\dA{D\bigl(\A\bigr)}
\def\bor{\overline}

\def\inth{ \underline{\mbox{Hom}} }

\begin{document}

\begin{titlepage}

\begin{center}

\section*{Mapping Class Group Actions on Quantum Doubles}

\bigskip
December 1993
\vspace*{1.5cm}
\bigskip

{\large Thomas Kerler\\
 \medskip
Department of Mathematics\\
Harvard University\\
Cambridge, MA, USA}\\
kerler@math.harvard.edu\\

\end{center}
\vspace*{3cm}

{\small Abstract:     We study representations of the mapping class group of
the punctured torus on the double of a finite dimensional possibly
non-semisimple Hopf algebra
that arise in the construction of universal, extended topological field
theories. We discuss how for doubles the degeneracy problem of TQFT's is
circumvented. We find compact formulae for
the ${\cal S}^{\pm 1}$-matrices using the canonical, non degenerate forms of
Hopf algebras and the bicrossed structure of doubles rather than monodromy
matrices. A rigorous proof of the modular relations and the computation of
the projective phases is supplied using Radford's relations  between the
canonical forms and the moduli of integrals. We analyze the projective $SL(2,
Z)$-action on the center of $U_q(sl_2)$ for $q$ an $l=2m+1$-st root of unity.
It
appears that the $3m+1$-dimensional representation decomposes into an
$m+1$-dimensional finite representation and a $2m$-dimensional, irreducible
representation. The latter is the tensor product of the two
dimensional, standard representation of $SL(2, Z)$ and the finite,
$m$-dimensional representation, obtained from the truncated TQFT of the
semisimplified representation category of  $U_q(sl_2)\,$.}

\end{titlepage}

\newpage

\setcounter{chapter}{1}
\section*{1.Introduction}

Since the seminal paper of Atiyah [A] on the abstract definition of a
topological quantum field theory (TQFT)  much progress has been made in finding
non trivial examples and
extended structures. The most interesting developments took place in three
dimensions where actual models of quantum field theory, like rational conformal
field theories and Chern-Simons theory led to the discovery of new invariants.
See [Cr] and [Wi].

\medskip
 In an attempt to counterpart these heuristic theories by
mathematically rigorous constructions the field theoretical machinery had
been replaced by quasitriangular Hopf algebras, or quantum groups. The
resulting invariants are described in  [TV] and [RT]. From here it is
not hard to understand  how to associate a TQFT  to a rigid, abelian,
monoidal category and an extended TQFT to a braided tensor category (BTC).
In order for these theories to be well defined  one has to make a few more
assumptions. One is that the category shall contain only a finite number
of inequivalent, simple objects, i.e., it is rational. The other is
a  technical non degeneracy condition, called ``modularity'' in [T],
 which is to assure that elementary cobordisms are associated to
identifications rather than projections. Alternatively, if the modularity
condition fails to hold, it is standard in the Atiyah [A] description to
define a truncated TQFT by reducing the vectorspaces to the images of the
projections.

\medskip
All of the mentioned TQFT's are semisimple, i.e., they rely on the
decompositions into simple objects. Clearly, semisimplicity cannot be
an assumption of fundamental but only of techncial nature. In seeking
{\em universal} constructions of TQFT's, which do not refer to decompositions,
one should thus not only generalize the existing ones to non semisimple
theories but also gain a deeper understanding of the structure underlying
 them.

\medskip
A  partial answer for the genus one case of how a universal TQFT should look
like had been given by Lyubachenkov in [Ly]. There representations of the
mapping class
group $\cal D$ of the punctured torus are constructed as a subgroup of  the
$End$ set of a coend in a BTC with certain finiteness conditions.
For the representation category of a finite dimensional Hopf algebra the
coend turns out to be the algebra acting on itself by the adjoint action.
A number of explicit formulae for the action of genus one mapping class
groups on Hopf algebras had been derived from this by Lyubachenkov and
Majid [LyM].

\medskip
One of the objectives of this paper is to give  natural definitions of the
modular operators and independent, rigorous proofs of the relations that
rely mainly on the theory of integrals on Hopf algebras as developed by Larson,
Sweedler and Radford. In doing so we will be able to give the precise relation
of the projective phases of the representation to the basic invariant of a
Hopf algebra obtained from the moduli.

\medskip
Starting with nonsemisimple Hopf algebras it is a natural question to ask
how the universal TQFT relates to the reduced TQFT defined by the
semisimplified
representation category of the same algebra. In the second part we
 give the precise connection for the mapping class group $\,SL(2,{\bf Z})\,$ of
the
closed torus and the quantum group $\,U_q(sl_2)\,$ . In the universal picture
the representation of $\,SL(2,{\bf Z})\,$ is found as  the restriction of the
action of $\cal D$ to the center. The usual modular
representation will appear in a tensorproduct with the fundamental,
algebraic
representation besides an additional, inequivalent finite representation.

\medskip
 In order to give an idea where these results fit into the general framework of
a TQFT we give here an outline of the  constructions of an extended
three dimensional TQFT with BTC's. The axioms are essentially due to Kazhdan
and Reshetikhin, [KR], and differ from other definition in that they make no
use of
higher algebraic structures like 2-categories. We shall give the objects
assigned to compact, oriented surfaces with boundaries both in the case of the
TQFT constructed in [RT] for semisimple
categories and for the universal TQFT associated to a Hopf algebra $\A\,$ .

\bigskip
\noindent
\underline{Extended Three Dimensional Topological Quantum Field
Theories\,:}

As in [A] an extended TQFT is defined as
a functor or, more precisely, a  collection of functors from cobordism
categories to abelian categories over an algebraically closed field $k$.

To a given one dimensional manifold $S$ we can associate a cobordism
category
$Cob_S$ as follows: The objects of the category are compact, oriented
two-folds $\Sigma$ with coordinate maps $S\,\isto\,\partial\Sigma $.
A morphisms between $\Sigma_1$ and  $\Sigma_2$ is a 3-fold $M$ whose
 boundary is parametrized by \mbox{$ \,-\Sigma_1\coprod_S\Sigma_2\isto\partial
M\,\,$}. The
composition of two morphisms is are given by an identification along a common
surface. An extended TQFT  assigns to every
surface $S$ a category ${\cal C}_S$  and a functor
$$
\Phi_S\,:\,Cob_S\,\longrightarrow\,{\cal C}_S\;.
$$
Assuming that ${\cal C}_{\emptyset}\,=\,Vect(k)\,$ this implies
 the original definition of [A]. We have a natural inclusion of
categories $Cob_S\times Cob_{S'}\,\hookrightarrow \,Cob_{S\amalg S'}$. For
the respective abelian categories we also assume a functor
\beq\label{union}
\odot\,:\,{\cal C}_S\times {\cal C}_{S'}\,\to\,{\cal C}_{S\amalg S'}\,
\eeq
compatible with $\Phi$. We require this to be a tensorproduct of abelian
categories in the sense of [D]. Note that this is consistent with ${\cal
C}_{\emptyset}\odot{\cal C}=Vect(k)\odot{\cal C}\cong {\cal C}\,$.

\medskip
A standard consequence of this are representation of mapping class groups.
To see this we consider $M\,=\,\Sigma\times
{\rm I}\coprod_{\sim}\partial\Sigma\,$ where  the relation $\sim$ is $(s,t)\sim
s\;\forall s\in\partial\Sigma,\;t\in {\rm I}$ and I is the unit interval.
For the boundary $\partial
M=\Sigma\coprod_{\partial\Sigma}\Sigma$ we choose different coordinate maps
for the two boundary pieces coinciding on $\partial\Sigma$. If we denote
by $\diff\bigl(\Sigma,\partial\Sigma\bigr)$ the group of homeomorphisms of
$\Sigma$ to itself which are identity on the boundary we obtain from these
cobordisms a representation:
\beq\label{R1diff}
\pi_o\Bigl(\diff\bigl(\Sigma,\partial\Sigma\bigr)\Bigr)\,\longrightarrow\,End_{{\cal
C}_{\partial\Sigma}}\Bigl(X_{\Sigma}\Bigr)\;.
\eeq
Here we denoted by
$X_{\Sigma}\,=\,\Phi_{\partial\Sigma}\Bigl(\bigl(\Sigma,\partial\Sigma\bigr)\Bigr)\,$.

\medskip
Next we formulate the axiom that leads to lower dimensional cobordism
functors. To this end suppose that $S=A\coprod B$ and $S'=B\coprod C$ then for
tensor categories
the contraction functor $\,Hom(1,\_\otimes\_ )\,:{\cal C}_B \times {\cal C}_B
\to Vect(k)\,$ induces a bilinear, covariant functor
\beq\label{Cfun}
{\cal C}_{A\amalg B}\times {\cal C}_{B\amalg C}\,\to {\cal C}_{A\amalg C}\;.
\eeq
On the side of the cobordism categories we consider two three manifolds
$M$ and $M'$
that belong to $Cob_{A\amalg B}$ and $Cob_{B\amalg C}$  respectively. We
can consider half tubular neighborhoods of the
1-folds $B$ in the boundaries of $M$ and $M'$. These define oriented
ribbon graphs in the boundaries along which we can glue the two manifolds
$M$ and $M'\,$. The result is again a three manifold $M\amalg_B M'$. The
boundary pieces are the  boundary pieces of the individual 3-folds glued
along $B$. This way we obtain a cobordism in $Cob_{A\amalg C}$ from
$\Sigma_1\coprod_B\Sigma_1'$ to $\Sigma_2\coprod_B\Sigma_2'$. The
assignment
\beq\label{cobfun}
Cob_{A\amalg B}\times Cob_{B\amalg C}\,\to\,Cob_{A\amalg C}
\eeq
is easily seen to be a functor. The next axiom of an extended TQFT
asserts that the functors $\Phi$ intertwines the two functors in
(\ref{Cfun}) and (\ref{cobfun}).

\medskip
This axiom allows us to define a functor from the category of 2-cobordisms
between 1-folds and the category of abelian, tensor categories.
The assignement of  morphisms is given by the composition:
\beq\label{dfuncs}
{\cal F}_{\Sigma}\,:\,{\cal C}_A\stackrel{1\odot X_{\Sigma}}{\hbox to
30pt{\rightarrowfill}}{\cal C}_A\odot{\cal C}_A\odot {\cal C}_B
\stackrel{Hom(1,\_\otimes\_\,)\odot id}{\hbox to
30pt{\rightarrowfill}}{\cal C}_B
\eeq
Here $\Sigma$ denotes a 2-manifold cobording the pieces $A$ and $B$ by some
coordinate maps \mbox{$-A\rightarrow\,\partial\Sigma\,\leftarrow B\,$}.

\medskip
In order to check functoriality of $A\to {\cal C}_A$ and $\Sigma \to {\cal
F}_{\Sigma}$ we consider again the manifold $M=\Sigma\times{\rm
I}\coprod_{\sim}\partial\Sigma$ as in (\ref{R1diff}) now with the same
coordinate maps for the boundary pieces but two components for
$\partial\Sigma$. Specializing to surfaces of the form
$\Sigma\,=\,S\times{\rm I}\,$ we get as in (\ref{R1diff}) a  homomorphism
\beq\label{diffS}
\pi_o\Bigl(\diff\bigl(S\bigr)\Bigr)\,\longrightarrow End_{Cat}\bigl({\cal
C}_S\bigr)
\eeq
For compact $S$  and by (\ref{union}) we easily identify (\ref{diffS}) as
the homomorphism from the permutation group of circles and to the
permutations of tensor factors.

\medskip
The functors associated to the elementary cobordisms, given by  spheres with
one, two, and three
punctures (denoted $P_1\,$, $P_2\,$ and $P_3$ respectively)
have a specific meaning for the circle category. Since $P_2$, seen as a
cobordism from $S^1$ to $S^1$
with the same coordinate maps, is a unit in the cobordism category we want
the associated ${\cal F}_{P_2}$ to be the identity functor in the basic
category ${\cal C}_1$ of the circle. Regarding $P_3$ as a cobordism from
$S^1\amalg S^1$ to $S^1$ the associated functor defines a tensor product
${\cal F}_{P_3}=\otimes\,:\,{\cal C}_1\odot{\cal C}_1\to {\cal C}_1\,$,
which we assume to be same as the one used in (\ref{dfuncs}).
Finally, the functor of $P_1:\emptyset \to S^1$ clearly gives the injection
of an identity object with respect to $\otimes$ and $P_1:S^1\to \emptyset$ is
assigned to
the invariance functor $Hom(1,\_\,)\,$.

\medskip
In an  extended TQFT we can also consider 3-cobordisms of 2-cobordisms,
which yield natural transformations. More precisely, let $M$ have boundary
pieces $\Sigma_i$, $i=1,2\,$ and $\partial\Sigma_i=A\coprod B\,$.
The functor $\Phi_{\partial\Sigma}$ associates to the surfaces $\Sigma_i$
objects $ X_i \in {\cal C}_A\odot{\cal C}_B\,$ and a morphism $f_M\in
Hom\bigl(X_1,X_2\bigr)\,$. For an object $Y\in {\cal C}_A$ we apply to
the morphism $id\odot f_M\,:\,Y\odot X_1\to Y\odot X_2\,$ the functor
$Hom(1,\_\cot\_\,)\odot id\,$ as in (\ref{dfuncs}) to give us a morphism
$\tilde f_M(Y)\,:\,{\cal F}_{\Sigma_1}(Y)\to{\cal F}_{\Sigma_2}(Y)\,$.
 It is easy to see that this defines a natural transformation
 $\tilde f_M\,:\,{\cal F}_{\Sigma_1}\stackrel{\bdot}{\to}{\cal
F}_{\Sigma_2}\,$ and thereby a functor
$\,Cob_{A\amalg B}\to Funct\bigl({\cal C}_A,\,{\cal C}_B\bigr)\,$.

\medskip
A special type of natural transformations are generated by cobordisms of
the form $M=S\coprod_{\alpha}\bigl(S\times{\rm I}\times{\rm
I}\bigr)\coprod_{\beta}S\,$ with relations $\alpha:s\sim (s,0,t)\,$ and
$\beta:(s,1,t)\sim F(s,t)\,\forall
s\in S\; t\in {\rm I}\,$. Here $F$ is a homotopy in the set of
homeomorphisms $\diff (S)$ of $S$ to itself. Confining ourselves to loops,
i.e.,
$F(s,1)=F(s,0)=s\,$, we obtain a homomorphism
\beq\label{pi1diff}
\pi_1\bigl(\diff(S)\bigr)\,\to\, Nat\bigl(id,\,id\bigr)
\eeq

\medskip
Reconsidering the elementary cobordisms $P_i\,$, we can discuss some elementary
natural transformations that identify the circle category ${\cal C}_1$ as a
BTC. The $2\pi$ rotation  of $S^1$ generating $\pi_1\bigl(\diff(S^1)\bigr)$
gives us by (\ref{pi1diff}) a natural transformation, denoted $\theta\in
Nat(id)\,$. We can also cobord the surface $P_3$ to $P_3$ with exchanged
coordinate maps for the $S^1\amalg S^1$ piece of the boundary by moving the
circles around each other in one of two directions. The TQFT assigns a
transformation $\epsilon^{\pm}\in Nat\bigl(\otimes,\,P\otimes\bigr)\,$.
 The square of this cobordism is homeomorphic to to the one where annuli
around the punctures are twisted by $2\pi$ so we obtain the identity of
natural transformations:
\beq\label{topbal}
\epsilon(Y,X)\epsilon(X,Y)\,=\,\theta(X\otimes
Y)\,\theta(X)^{-1}\otimes\theta(Y)^{-1}
\eeq
This means $\theta$ is a balancing of ${\cal C}_1$. The associativity
constraint is obtained in a similar way.

\medskip
Let us discuss for  a  surface $\Sigma'$ whose boundary is the union
on $n$ circles and the corresponding closed surface $\Sigma$ a connection
between (\ref{R1diff}) and (\ref{diffS}). We have fibrations
$$
\diff\bigl(\Sigma',\partial\Sigma'\bigr)\,\hookrightarrow
\diff\bigl(\Sigma'\bigr)\twoheadrightarrow \diff\bigl(\partial\Sigma'\bigr)
$$
$$
{\rm and}\qquad\qquad\qquad
\diff\bigl(\Sigma'\bigr)\,\hookrightarrow\,\diff\bigl(\Sigma\bigr)\twoheadrightarrow K_n\qquad\qquad\qquad
$$
where $K_n$ is the symmetrized configuration space of $n$ points in
$\Sigma\,$. From the long exact sequence for the first fibration and the
injection of the second we obtain the top row of the following commutative
diagram:
\beq
\bar{ccccc}\label{fibcom}
\pi_1\Bigl(\diff\bigl(\partial\Sigma'\bigr)\Bigr)\,&\longrightarrow
&\,\pi_0\Bigl(\diff\bigl(\Sigma',\partial\Sigma'\bigr)\Bigr)\,&\longrightarrow
&\,\pi_0\Bigl(\diff\bigl(\Sigma\bigr)\Bigr)\\
\Bigg\downarrow&&
\Bigg\downarrow&&
\Bigg\downarrow
\\
Nat_{{\cal C}_{\partial\Sigma'}}\Bigl(id,\,id\Bigr)\,&
\longrightarrow
&\,
End_{{\cal C}_{\partial\Sigma'}}\Bigl(X_{\Sigma'}\Bigr)\,
&\,
\longrightarrow
&\;
End_k\Bigl(V\bigl(\Sigma\bigr)\Bigr)
\ear
\eeq
In the bottom row the left map is simply the evaluation of a natural
transformation on an object. The second homomorphism is given by the
invariance functor $\,Hom(1,\_\,)^{\odot n}\,$ acting on ${\cal
C}_{\partial\Sigma'}\,\cong\,{\cal C}_1^{\odot n}\,$ and
$V\bigl(\Sigma\bigr)\,=\,Hom(1,\_\,)^{\odot n}\bigl(X_{\Sigma'}\bigr)\,$
is the vectorspace associated to the closed surface.

\bigskip
\underline{Examples and the Degeneracy Problem :} The objects associated to
punctured surfaces can be identified up to isomorphie for two types of
categories. One is a semisimple, rational BTC ${\cal C}_o\,$ with simple
objects ${\cal I}\,$ the other is the representation category $R({\cal A})$
 of a finite dimensional Hopf algebra ${\cal A}\,$. Quite generally it
is possible to produce a semisimple, rational category from $R({\cal A})$
by a generalized GNS construction with respect to a canonical categorial
trace $tr\,$, see for example [K]. Thus in principle there are two ways of
constructing
TQFT's  from a given Hopf algebra $\cal A$ which will lead to different
representations, e.g., of mapping class groups. The precise connection in
one example will be discussed in an example in the last chapter.

\medskip
The assignement of objects for the two punctured sphere is easily inferred
from ${\cal F}_{P_2}=id\,$  and formula (\ref{dfuncs}). In ${\cal C}_o$
the answer is  $X_{P_2}\cong\sum_{j\in{\cal I}}j\odot j^{\vee}\,$ and
in $R({\cal A})$ the module $X_{P_2}$ is given by $\cal A$ with ${\cal
A}^{\odot 2}\,$ -action aiven by $a\odot b.(x)\,=\,axS(b)\,$. Moreover,
${\cal F}_{P_3}=\otimes\,$ implies that $\,X_{P_3}\cong \sum_{ij\in {\cal
I}}
i\odot j\odot (i\otimes j)^{\vee}\,$ or $\,X_{P_3}\,=\,{\cal A}\otimes{\cal
A}$ with ${\cal A}^{\odot 3}$-action $\,(a\odot b\odot c).(x\otimes
y)\,=\,(ax\otimes by)\Delta\bigl(S(c)\bigr)\,$. This allows us to identify
 the object associated to the punctured torus $T'$  with $\partial T'=
S^1\,$ by contracting the objects $X_{P_3}\,$ and $X_{P_2}\,$ along the
category of the $S^1\amalg S^1\,$-boundary pieces. In ${\cal C}_o$ the
resulting
object is $X_{T'}\,=\,\sum_{j\in{\cal I}}j\otimes j^{\vee}\,$ and in
$R({\cal A})$ by the module $\cal A$ with adjoint action.
The objects of all other surfaces are now found easily by sewing along
circles. For example the surface $\Sigma_{g,1}$ of genus $g$ with one
puncture is assigned to $X_{T'}^{\otimes g}\,$. The object of the
$(n+1)$-punctured sphere $P_{n+1}$ has object $X_{P_{n+1}}\,=\,\sum_{i_k}
i_1\odot\ldots\odot i_n \odot (i_1\otimes\ldots\otimes i_n)^{\vee}$ in
${\cal C}_o$ and in $R({\cal A})\,$ the module $X_{P_{n+1}}=\,{\cal
A}^{\otimes n}$ where the ${\cal A}^{\odot(n+1)}$-action is  given by the
obvious generalization of the cases $n=2,3\,$. The object for a  general
compact, orientable surface is found by sewing $X_{P_{n+1}}$ and
$X_{\Sigma_{g,1}}\,$.
For $R({\cal A})$ this gives for example the module
 $Hom_{\cal A}({\cal A}^{\otimes g},\,{\cal A}^{\otimes n})$ of
intertwiners for one of the $\cal A$-actions.

\medskip
Let us discuss the case $g=1,\,n=1$ in some more detail. The mapping class
group ${\cal D}\,=\,\pi_o\Bigl(\diff\bigl(T',\partial T'\bigr)\Bigr)\,$
maps by (\ref{R1diff}) into  $\,End\bigl(T'\bigr)\,$ so that we obtain in
$R({\cal A})$ an action of $\cal D$ on $\cal A$ intertwining the adjoint
action.  Following (\ref{fibcom}) we obtain a representation of the modular
group $\pi_o\bigl(\diff(T)\bigr)\,$ on $V(T)=Hom(1,T)$, which for
$R({\cal A})$ is just the restriction of the $\cal D$ action to the center
$Z({\cal A})=Hom(1,{\cal A})\,$. In order to  interpret the rest  of
(\ref{fibcom}) recall that for $R({\cal A})$ the natural transformations of
the identity functor are given by the action of central elements of $\cal
A\,$. In particular  the generator  $\theta$ of
$\pi_1\bigl(\diff(S)\bigr)\,$ acts on $\cal A$ as $ad(v)$ where
$v=\theta({\cal A})$ is the central ``ribbon element'', see [RT]. The Dehn
twist along the boundary can also be given by ${\cal S}^4$ where $\cal S$
is the standard generator of $\cal D\,$. The restriction of ${\cal
S}^4=ad(v)$ to the center is clearly trivial. The second generator of
$\cal D$, the Dehn twist at a handle, $\cal T$ is given by the action of
$\theta$ on the constituent $X_{P2}$, i.e., by multiplication of $v$ on $\cal
A\,$.

\medskip
The definition of a TQFT we presented so far is not quite complete.
Clearly, there are many ways of sewing up a surface $\Sigma$ so we have
many ways to construct the object $X_{\Sigma}\,$. For example instead of
using the center of $\cal A$ as the vectorspace for
the closed torus $V(T)$ we can also choose the space $Hom({\cal A},1)$ -
which is isomorphic to the space of characters on $\cal A$ - or we
could have chosen the endomorphism set
$End\bigl(X_{P_2}\bigr)\,=\,End_{{\cal A}^{\odot 2}}\bigl({\cal
A}\bigr)\,$. These spaces are isomorphic to each other but there is no one
canonical isomorphism identifying two of them.
Instead the sewing procedure used to find the object defines a surface
with a cut diagram or decoration. Thus we should take as objects of the
cobordism categories surfaces $\Sigma$ together with a Lagrangian subspace of
$H_1(\Sigma, {\bf R})$ which nust be compatible with the cobording 3-manifolds.
The functor of the TQFT is now allowed to have projective phases. This means
for two cobordisms $M_1$ and $M_2$ with a common, decorated  boundary component
that
\beq\label{projfun}
\Phi(M_1M_2)\,=\,c^{\mu}\Phi(M_1)\Phi(M_2)
\eeq
where $\mu\,$ is the Maslov index of a triple of Lagrangian subspaces defined
by the cobordisms. It also measures the non-additivity of the signature of the
4-manifolds cobording the $M_i$ to the corresponding union of handlebodies.
If the $M_i$ are invertible morphisms in the cobordism category we obtain
projective representations of the modular groups. For details see [T].
 The main result of the first chapter is the relation
of the phase $c\,$ to intrinsic invariants of the Hopf algebra $\cal A\,$.

\medskip
In order to discuss the modularity condition we recall how the  $\cal S$
matrix can be obtained from the [RT]-construction for standard TQFT's with
$S=\emptyset\,$. The cobordism describing the action of $\cal S$ on
$T'$ is a 3-manifold whose boundary is $\partial M =T'\coprod_{S^1}T'\,$,
the closed surface of genus two, and can thus be considered a cobordism
$\Sigma_2\to\emptyset\,$. In [RT] the vectorspace associated to $\Sigma_2$
is
$\bigoplus_{ij} Hom\bigl(i\otimes i^{\vee}\otimes j\otimes
j^{\vee},\,1\bigr)\,$. The linear form assigned to $\Sigma_2\to\emptyset$ is
found by computing the the invariant in $S^3$ of the  ribbon graph
embedded in the outside of $\Sigma_2\,$. On a vector $f$ its value is
$$
1\,\stackrel{coev\otimes coev}{\hbox to 50pt{\rightarrowfill}}\,i\otimes
i^{\vee}\otimes j\otimes j^{\vee}\,\stackrel{1\otimes\epsilon^2\otimes
1}{\hbox to 40pt{\rightarrowfill}}\,i\otimes i^{\vee}\otimes j\otimes
j^{\vee}\,
\stackrel {f}{\hbox to 30pt{\rightarrowfill}}\,1
$$
In the description of  an extended TQFT we have to consider this as the
matrix element of ${\cal
S}\,\in\,End(T')\,\cong\,\bigoplus_{ij}Hom\bigl(j\otimes
j^{\vee},\,i\otimes i^{\vee}\bigr)\,$. Thus on a summand we have
$$
{\cal S}\,:\,j\otimes j^{\vee}\,\stackrel { coev \otimes 1}{\hbox to
25pt{\rightarrowfill}}i\otimes i^{\vee}\otimes j\otimes
j^{\vee}\,\stackrel{1\otimes \epsilon^2 \otimes 1}{\hbox to
40pt{\rightarrowfill}}\,
i\otimes i^{\vee}\otimes j \otimes j^{\vee}\,\stackrel {1\otimes c_j}{\hbox
to 30pt {\rightarrowfill}}\, i\otimes i^{\vee}
$$
where $c_j$ is proportional to $j\otimes j^{\vee}\stackrel
{\epsilon}{\longrightarrow}\, j^{\vee}\otimes j\,\stackrel
{ev}{\longrightarrow}\,1\,$. The generalization of this formula to non
semisimple categories is described by [Ly] and will be reviewed in the next
chapter.

The matrix elements of the restriction of $\cal S$ to
$V(T)=Hom\bigl(1,X_{T'}\bigr)=\bigoplus_jHom(1,j\otimes j^{\vee})\cong
k^{\cal I}$ are given by ${\cal S}_{ij}\,=\,tr_{i\otimes
j}\bigl(\epsilon(i,j)\epsilon(j,i)\bigr)\,$ where $tr$ is the usual trace
of a balanced category.

\medskip
A priori the operations $\cal S$ and $\cal T$ defined for a general
semisimple ${\cal C}_o$ do not yield a projective representation of
$SL(2,Z)\,$ unless we impose one further condition. This is the
rather specialized ``modularity condition'' introduced in [T] asserting
 that the $\cal S$- matrix is invertible.
 In case this condition is violated we may still apply Atiyah's
prescription  and reduce  the space $k^{\cal I}$ by the projection $P={\cal
S}{\cal S}^-$ where the matrix ${\cal S}^-_{ij}={\cal S}_{ij^{\vee}}$ is
assigned by the [RT]- prescription to the inverse cobordism.

\medskip
For example if ${\cal C}_o$ is a symmetric category the $\cal S$ matrix is
of rank one so the $SL(2,{\bf Z})$ representation is one dimensional.  A
degeneracy  problem occurs quite generally if $\cal I$ contains a subset${\cal
I}_o$,
of irreducible objects, which braid trivially, i.e.,
$\epsilon(k,j)\epsilon(j,k)=1$ for all $k\in{\cal I}_o,\;j\in{\cal I}\,$.
In case ${\cal I}_o$ is a subgroup of invertibles $\{\sigma\}$ we have for the
natural action  of its elements on $k^{\cal I}$ that
${\cal S}\sigma={\cal S}\,$. Hence $\cal S$ and $\cal T$ can be defined on
the orbit space $im\bigl(\sum_{\sigma\in{\cal I_o}}\sigma\bigr)\,$, where
we can hope for the modularity condition to hold.

\medskip
This situation occurs for the semisimplified representation categories
of quantum groups at certain roots of unity.  The example we will come back
to in the last chapter  is $U_q\bigl(sl_2\bigr)\,$ where $q^{1/2}$ is an
$l=2m+1\,$- th root of unity. We have $|{\cal I}|=2m$ and the
 $2m$-th representation braids trivially and is invertible of order two.
The truncated theory yields an $m$-dimensional representation of $SL(2,{\bf
Z})\,$.

\medskip
The problem of degeneracy is resolved in a very natural way in the
universal picture for $R({\cal A})$ by choosing  $\cal A$ to be a double
constructed algebra. In this situation we find very simple formulae for
$\cal S$ and its inverse.

\bigskip
\noindent
\section*{ Survey of Contents and Summary of Results: }

 In Chapter 2 we define and
study the action of operators generating the mapping class group ${\cal D}\,:=$
 $\pi_o\bigl( \diff (T,D)\bigr)$
on the double $\dA$ of a finite dimensional Hopf algebra. We start in Section 1
with a review of the bicrossed structure of a double and properties of an
isomorphism $\dA^*\isto\dA\,$. These are in particular the relations between
traces and  characters on $\dA$ and central and group like elements in $\dA\,$.
 We also recall the definitions of canonical and balancing elements in
quasitriangular Hopf algebras. For later application we derive a relation
for the monodromy matrix of $\dA\,$. The next section is a recollection from
[Rd] of relations between several non degenerate bilinear forms and moduli
defined by the integrals of a finite dimensional Hopf algebra. This
leads for $\dA$  to the Drinfeld-Radford formula $S^4=Ad(g)$.
In Section 3 we determine the integral and cointegral of a double $\dA\,$.
In particular we find that the comodulus is trivial and that the modulus is the
canonical element $g\,$. This allows us to show that a pair of non-degenerate,
canonical traces on $\dA$ can be defined very simply from the natural
contraction on $\dA\,$. The balancing of a double $\dA\,$ is related in Section
4
to second order roots of the moduli and a fourth order root $\nu$ of the
$\omega$-
invariant of $\A\,$. Guided by categorial constructions in [Ly] we define in
Section 5 the action of the generators of the mapping class group $\cal D$
on $\dA\,$. We obtain an intriguingly simple expression for the actions of
$\cal S$ and ${\cal S}^{-1}\,$  involving only the non degenerate forms
on $\A$ and $\A^*$ and the bicrossed isomorphism of $\dA\,$. Similarly we have
a formula for the braided antipode. The results from all previous sections are
used in Section 6 to give a rigorous proof of the modular relations and
determine the projective phase of the universal TQFT as $\nu^{-3}\,$.

\medskip
In Chapter 3 we find the structure of the representation of $SL(2,{\bf Z})$ on
the center of $U_q\bigl(sl_2)\,$ by restricting the action of $\cal D\,$.
The non-degenerate forms and moduli of the double of $B_q$, the Borel algebra
of $U_q\bigl(sl_2)\,$, are given in Section 1. In Section 2 we determine the
center of $D(B_q)$, which for the TQFT is the vectorspace of the torus. If $q$
is an
$l=2m+1$-st primitive root of unity it is given by ${\bf C}[{\bf Z}/l]\otimes
{\cal V}\,$. Here ${\cal V}$ is a $3m+1$-dimensional algebra with a basis of
$m+1$ idempotents and $2m$ nilpotents. The balancing element of $D(B_q)$ is
expressed in terms of this basis, see Section 3. In doing so we propose a
method to generate new partition identities.
In Section 4 we compute the matrix elements of the $SL(2,{\bf Z})$-action on
the center of $D(B_q)\,$. This requires us to find transformations from
 the PBW basis of $U_q\bigl(sl_2)\,$ to the algebra $\cal V\,\otimes{\bf C}[ K
]\,$, where $K$ is a Cartan element. We analyze this representation in Section
6 . We find
evidence for the decomposition of the representation into two irreducibles.
One of which is a finite, $m+1$ - dimensional representation, the other is the
tensor product of the two dimensional standard representation of $SL(2,{\bf
Z})$
and the finite, m-dimensional representation obtained from the semisimplified
representation category of   $U_q\bigl(sl_2)\,$. The conjecture is verified in
the last section. Here we find the decomposition and the explicit finite
representations for  two  non trivial roots of unity.

\section*{Acknowledgements } I thank D. Kazhdan for many, very useful
discussions and S. Majid for discussion and help with the references. I
thank H. Wenzl and the Department of Mathematics of UCSD, where part of this
work was done, for kind hospitality. This work was supported in part by
NSF grant  DMS-9305715.

\newpage

\setcounter{chapter}{2}
\section*{2. Mapping Class Group Action on Doubles}
\bigskip
\ub{1.) Double Algebras and Balancing :}
In this section we recall some basic facts and notions on Hopf algebras that we
will use later.
For a finite dimensional Hopf algebra $\A$ over a field $k$ we denote by $\oA$
the dual Hopf algebra with opposite comultiplication. We shall always assume
that $\A$ has a counit $\epsilon$
and an invertible antipode $S\,$. The antipode of $\oA$ is thus given by
${S^{-1} }^*\,$.

For $\lambda , \mu\,\in\,\oA$ and $a,b\,\in\,\A$ we have the following
relations:
\bea
<\lambda\mu,a> =<\lambda\cot \mu, \Delta(a)>\qquad&&\quad <\Delta(\lambda),ab>
=<\lambda, b\cot a>\label{dual1}\\
\Delta\cot id(R) =R^{13}R^{23}\qquad \qquad &&\quad\qquad id \cot \Delta (R)
=R^{13}R^{12}\label{dual2}\\
<S^{\mp 1}(l),a> =<l,S^{\pm 1}(a)>\qquad&& \;S\cot id (R)=id\cot
S^{-1}(R)=R^{-1}\label{dual3}
\eea
Here $<,>:\oA\cot\A\to k$ is the usual contraction and $R$ is the canonical
element $R\in \A\cot\oA\,$. Thus, if  $\{e_i\}$ is a basis of  $\A$ and
$\{f_i\}\,$ the respective dual basis of  $\oA$  we can write  $R=\sum_ie_i\cot
f_i\,$.
A bicrossed product of the two algebras is a Hopf algebra $D$ which contains
$\A$ and $\oA$ as sub-Hopf algebras such that  $\bdot \,:\,\A\cot\oA \isto D\,$
is an isomorphism, where $\bdot$ is the multiplication in $D\,$. Clearly, a
bicrossed structure is uniquely determined by an isomorphism
$\bowtie:\A\cot\oA\isto\oA\cot\A$ which by $\bdot\bowtie=\bdot$ defines an
associative product, such that the coproduct on $D$ defined by the coproducts
on
$\A$ and $\oA$ extends to a homomorphism into $D^{\otimes 2}$. In [Dr0] it is
shown that there is precisely one bicrossed product $\dA$, the double,  such
that
\beq\label{rcom}
R\Delta(y)\,=\,\Delta'(y) R \qquad {\rm for\; all} \quad y\in\dA\;.
\eeq
Here $\Delta'=\tau\circ\Delta$ is the opposite comultiplication and $\tau$ is
the flip $\tau(a\cot b)=b\cot a\,$. The bicrossed structure is given explicitly
by
\bea\label{bic+}
\bowtie\,:&\,\A\cot\oA
\stackrel{\tau}{\hbox to 15pt{\rightarrowfill}}
\oA\cot\A
\stackrel{\Delta\cot\Delta'}{\hbox to 30pt{\rightarrowfill}}
\oA\cot\oA\cot\A\cot\A
\stackrel{ 1\cot S\cot 1\cot 1}{\hbox to 40pt{\rightarrowfill}}
\oA\cot\oA\cot\A\cot\A\nonumber\\
&\stackrel{ 1\cot <,>\cot 1}{\hbox to 45pt{\rightarrowfill}}
\oA\cot\A
\stackrel{\Delta'\cot\Delta}{\hbox to 30pt{\rightarrowfill}}
\oA\cot\oA\cot\A\cot\A
\stackrel{ 1\cot <,>\cot 1}{\hbox to 45pt{\rightarrowfill}}
\oA\cot\A
\eea
If we use the usual abbreviation $\Delta(a)=\sum_i a_i'\cot a'_i = a'\cot
a''\,$, $\Delta^2(a)=a'\otimes a''\otimes a'''$ (\ref{bic+}) can be summarized
in the formula
$$
\bowtie(y\otimes \lambda)\,=\,\lambda''\otimes
y''\bigl<\lambda',y'\bigr>\bigl<S(\lambda'''),y'''\bigr>
$$
The inverse is given similarly by
\bea\label{bic-}
\bowtie^{-1}\,:&\,\oA\cot\A
\stackrel{\Delta'\cot\Delta}{\hbox to 30pt{\rightarrowfill}}
\oA\cot\oA\cot\A\cot\A
\stackrel{ 1\cot S\cot 1\cot 1}{\hbox to 40pt{\rightarrowfill}}
\oA\cot\oA\cot\A\cot\A
\stackrel{ 1\cot <,>\cot 1}{\hbox to 45pt{\rightarrowfill}}
\oA\cot\A\nonumber\\
&\stackrel{\Delta\cot\Delta'}{\hbox to 30pt{\rightarrowfill}}
\oA\cot\oA\cot\A\cot\A
\stackrel{ 1\cot <,>\cot 1}{\hbox to 45pt{\rightarrowfill}}
\oA\cot\A
\stackrel{\tau}{\hbox to 15pt{\rightarrowfill}}
\A\cot\oA
\eea
For a Hopf algebra $D\,$, with $dim(D)<\infty$  we denote by  $G(D)$  the
finite group of {\em group like} elements $g$, characterized by
$\Delta(g)=g\cot g\,$. Also we shall use the notation  $Ch(D)\,\cong\,G(D^*)$
for the group of one dimensional representations of $D\,$.
For doubles we have the following easy fact:
\blm\label{dougrou}
For a Hopf algebra $\A$ the multiplication map $\bdot:\A\cot\oA \isto\dA\,$
yields
a group isomorphism:
$$
G(\A)\,\oplus\,Ch(\A)\,\isto G(\dA)\quad.
$$
Similarly the sum of restrictions yields:
$$
Ch(\dA)\isto  Ch(\A)\,\oplus\, G(\A)\quad.
$$
\elm
{\em Proof:}
For $b\in\dA$  let $\ub b$ be the corresponding element in  $End(\A)\,$.
For the coproduct this means $\ub{\Delta(b)}(x\cot y)\,=\,\Delta\bigl(\ub
b(yx)\bigr)$ which for $b\in G(\A)$ has to equal $\ub b (x)\otimes\ub b(y)\,$.
Inserting $y=1$ and
applying $\epsilon\otimes 1$ we find that $\ub a$ is of rank one and
$a=g\bdot\gamma$
where $g=\ub a(1)$ and $\gamma=\epsilon\circ\ub a\,$. Inserting instead $x=1$
and $y=1$
(applying $\epsilon\otimes\epsilon\,$) shows that $g\in G(\A)$ ($\gamma \in
Ch(\A)\,$).
 The adjoint action of $Ch(\A)$ on  the double $\dA$ stabilizes $\A$ and,
there, coincides
with the coadjoint action, i.e.,we have $\gamma\bdot y \bdot
\gamma^{-1}=\gamma\lcoa y \rcoa \gamma^{-1}\,$ for all $y\in\A\,$. Since the
coadjoint action on group likes in $\A$ is trivial the images of
$G(\A)$ and $Ch(\A)$ centralize each other and the inclusion factors into the
direct sum. Injectivity now follows from linear independence of group likes,
see [Ab], and injectivity of~$\bdot\,$.\hfill$\Box$

Here we used the notation $\lcoa$ ( $\rcoa$ ) as in [Ab] for the left (right)
action
of $D^*$ on a Hopf algebra $D$. Similarly, we use $(a\lac
\lambda)(y):=\lambda(ya)$
 for the left action of $D$ on $D^*$ and $\rac$ for the corresponding right
action.
 We also use the adjoint actions of $D$ on
itself given by $\,ad(a)(y)=a'yS(a'')\,$ and on $D^*$ given by
$\,ad_*(a)(\lambda)=a''\lac\lambda\rac S(a')$. The invariance in $D$ under the
adjoint action is precisely the {\em center} $Z(D)$ and the invariance  in
$D^*$ are the {\em $q$-characters } $\bor{C}(D)=\{\lambda \in D^*:
\lambda(xy)=\lambda(S^2(y)x)\}$, which were introduced in [Dr1]. In [Dr1]
it is shown that these two spaces are related to each other by the map
\beq\label{deff}
\bor{f} :D^*\to D\,:\,\lambda\to\lambda\otimes 1(M)\;.
\eeq
Here $M\in D^{\otimes 2}$ is the element
$M=\tau(R)R=\sum_{ij}f_je_i\cot e_j f_i=\sum_k m_k\cot n_k\;$.
In the case of a double $D=\dA$ $\{m_k\}$ and $\{n_k\}$ are differents basis of
$\dA$ so that $M$ is nondegenerate. The following is a slightly extended
version of a lemma in [Dr1].
\blm \label{veup}

\ben
\item The map $\bor{f}:\dA^*\isto \dA $ is an isomorphism of $\dA$-modules with
respect to the adjoint actions.
\item $\bor{f}:\bor{C}(\dA)\isto Z(\dA)$ is an isomorphism of algebras.
\item $\bor{f}:Ch(\dA)\isto G(\dA)$ is the group isomorphism $(g,\gamma
)\mapsto
(\gamma , g)$
\item We have $\bor{f}^*\circ S^*=S^{-1}\circ\bor{f}$
\een
\elm
{\em Proof}: The fact that $\bor{f}$ intertwines the actions of $\dA$ follows
from basic Hopf algebra relations, (\ref{rcom}) and the identity
\mbox{$\;\bigl(S(y')\cot 1\bigr)\, M \bigl(y'' \cot 1\bigr)\,  =
$}\mbox{$\;\bigl(S(y')\cot 1\bigr)\, M\, \bigl(y''\cot \bigl(
y'''S(y'''')\bigr)\bigr)\,$}
 \mbox{$=\,\bigl(S(y')\cot 1\bigr)\bigr( y''\cot y'''\bigl)\,M \,\bigl(1\cot
S(y'''')\bigr)\,=\,\bigl(1 \cot y' \bigr)\,M\,\bigr( 1\cot S(y'')\bigr)\,$}.

It is clear that $\bor{f}$ is an isomorphism. In particular we can write it as
the composition:
$$
\dA^*\stackrel{(\bdot\bowtie)^*}{\isto}\bigl(\A^*\cot \A\bigr)^*
=
\A\cot\A^*\,
\stackrel{\bdot}{\isto}\dA
$$
Clearly, the invariances are mapped isomorphically to each other
and a computation in [Dr1] shows that this is a homomorphism. In fact we have
$ \bor{f}(\chi\lambda)=\bor{f}(\chi)\bor{f}(\lambda) $ for \mbox{ $\chi\in
\bor{C}(\dA)\,$} and any \mbox{$\lambda\in\dA^*\,$}. {\em 3.)} follows from
 Lemma \ref{dougrou} and the form of $M\,$. Finally, (\ref{dual3}) implies
$S\cot S(M)=\tau (M)$ and thereby {\em 4.)}.\hfill$\Box$

It follows from (\ref{dual2}) that the $R$ matrices satisfy the Yang Baxter
equation $ R^{12}R^{13}R^{23}$ $=R^{23}R^{13}R^{12}\,$. For later computations
of
modular relation we derive here an analogous
equation for the $M$ matrices.

\blm\label{Meq}
For $M$ and bases $\,\{e_i\},\,\{f_i\},\,\{n_k\},\,\{m_k\}$ as above we have
$$
\bigl(\tau(M)\cot1\bigr)\bigl(1\cot M\bigr)\,=\,\sum_{kj}
 n_k'f_j\cot m_k \cot S^{-1}(e_j'')n_k''e_j'
$$
or equivalently
$$1\cot \lambda \cot 1 \Bigl( \bigl(\tau(M)\cot1\bigr)\bigl(1\cot
M\bigr)\Bigr)\,=\,\sum_{ij}\bor{f}(\lambda)'f_if_j\cot
S^{-1}(e_j)\bor{f}(\lambda)''e_i\;.
$$
\elm
{\em Proof:} If we multiply $R$ matrices from the left and right to the Yang
Baxter equation and permute the first and third factor we obtain
$(R^{-1})^{31}(R^{-1})^{32}R^{21} R^{31}\,=\,R^{21}(R^{-1})^{32}\,$. Applying
$1\cot 1\cot S^{-1}$ to this equation and using (\ref{dual3}) we find
$$
\sum_{ij}f_i\cot e_if_j\cot e_j\,=\,\sum_{i,j,k,l}f_if_jf_k\cot f_le_j\cot
S^{-1}(e_k)e_le_i
$$
Multiplication with $R\cot 1$ from the left and $1\cot R$ from the right yields
\bea
\bigl(\tau(M)\cot1\bigr)\bigl(1\cot M\bigr)\,&\quad =&
\sum_{tsijkl}e_tf_if_jf_k\cot f_tf_l e_j e_s \cot
S^{-1}(e_k)e_le_if_s\nonumber\\
& ^{by(\ref{dual2})}=&\sum_{ijkl}  e_l'f_if_j''f_k \cot f_l e_j \cot
S^{-1}(e_k)e_l''e_if_j'  \nonumber\\
& ^{by(\ref{rcom})}=& \sum_{ijkl}e_l'f_j'f_if_k\cot f_le_j \cot
S^{-1}(e_k)e_l''f_j''e_i\;.\nonumber
\eea
The formulas follow now from $\Delta(n_{(lj)})=e_l'f_j'\cot e_l''f_j''$ and
again (\ref{dual2}). \hfill $\Box$

Let us also record here the canonical elements from [Dr1] and [Ly1]
implementing the square of the antipode. They are defined by
\beq\label{defu}
u\,:=\,\sum_i S(f_i)e_i\qquad\quad {\rm and }\qquad\quad
\hat u\,:=\,\sum_iS^2(e_i)f_i
\eeq
and satisfy the relations
\bea\label{relu}
S^2(y)\,&=\,u y u^{-1}\,=\,\hat u y \hat u^{-1}\,,  \qquad\quad
\hat u\,=\,S(u)^{-1}\nonumber\\
 \qquad {\rm and} \qquad\qquad  & M\,=\,u\cot u \Delta(u^{-1})\,=\,\hat u^{-1}
\cot \hat u^{-1} \Delta(\hat u)\qquad\qquad\qquad\qquad
\eea
{}From $u$ and $\hat u$ one has two further elements of a quasitriangular Hopf
algebra $D$ with special properties:
\beq\label{defg}
g:=u \hat u \;,\quad {\rm with } \quad   g \in G(D) \qquad {\rm and} \qquad
S^4(y) = gyg^{-1}
\eeq
\beq\label{defz}
z:=u{\hat u}^{-1}\;, \quad {\rm with} \quad z \in Z(D) \qquad {\rm and } \qquad
M^2=z\cot z \Delta(z^{-1})
\eeq
\bigskip

\noindent
\ub{2.) Integrals, Moduli and Radford's Relations :} We start this section with
a review of basic facts from Hopf algebra theory and a summary of the formulae
in [Rd], we will use in this paper.
The analysis of integrals of Hopf algebras in [LSw] is based on the fundamental
Theorem of Hopf modules. It asserts that a Hopf module $M$ of a Hopf algebra
$D$ is free in the sense that $M^{cov}\cot D \isto M$ is an isomorphism of Hopf
modules. Here
$D$ acts on itself by multiplication and comultiplication, $M^{cov}$ is the
the coinvariance of the coaction and ${\isto}$ is given by the left action on
$M\,$. It is instructive to apply this to the situation where $M=D^*$ with
actions $h.\lambda := \lambda\rac S(h)$ and coaction
$\delta(\lambda)=\lambda\cot 1 \Delta \in End(D)=D\cot D^*\,$. The
isomorphism $J\cot D \isto D^*$ then implies that $J=\{ \lambda : \lambda \cot
1 \Delta(y)=1\lambda(y)\}$ - the space of {\em right integrals} - is one
dimensional and every nonzero element induces a nondegenerate bilinear form.
Analogous statements are found if we use left actions or consider Hopf modules
of the dual algebra. Let us fix once and for all a left integral $\mu$ and a
left cointegral $x$ with the properties
\beq\label{lints}
1\cot \mu \Delta(h)=1\mu(h),\qquad hx=\epsilon(h)x,\quad {\rm and} \quad
\mu(x)=1\;.
\eeq
As in [Rd] we use notations for the following isomorphisms:
\bea
\beta_l,\,\beta_r\,:\,D\isto D^*\quad &{\rm with} \qquad \beta_l(h)=\mu \rac
h\quad {\rm and } \quad \beta_r(h)=h\lac\mu\label{defbet}\\
\bor\beta_l,\,\bor\beta_r\,:\,D^*\isto D \quad &{\rm with }\qquad
\bor\beta_l(\lambda)=x\rcoa\lambda\quad {\rm and } \quad
\bor\beta_r(\lambda)=\lambda\lcoa x\;.\label{defborbet}
\eea
They intertwine the right and left actions as in
\beq\label{intbet}
\beta_l(kh)=\beta_l(k)\rac h\qquad{\rm and}\qquad\beta_r(kh)=k\lac\beta_r(h)\;.
\eeq
It is obvious that $xh$ is again a left cointegral for any $h \in D$. Hence by
uniqueness of $x$, we find $\alpha \in Ch(D)$ and for the dual situation $a\in
G(D)$ such that
\beq\label{modrel}
\alpha(h)x=xh\qquad {\rm and } \qquad \mu\cot 1 \Delta(h)=a\mu(h)\;.
\eeq
Since $D$ is finite dimensional, both the {\em modulus} $\alpha$ and the {\em
comodulus } $a$ are of finite order and
\beq
\omega:=\alpha(a)
\eeq
is a root of unity. Note, that in the following we use the opposite
comultiplication for $D^*$ so that, e.g., $S^{-1}=S^*\,(=\gamma \;{\rm in \
[Rd]})\,$. The antipode acts on the integrals as follows:
\bea\label{Sint}
S^{-1}(\mu)=\,a\lac\mu=&\omega \,S(\mu)=\,\omega\,\mu\rac a\\
S(x)=\alpha\lcoa x = &\omega S^{-1}(x)=\omega x \rcoa \alpha
\eea
The compositions of isomorphisms in (\ref{defbet}) and (\ref{defborbet}) are
given by the following formulae. Each one can be given on $D^*$ or the adjoint
one on $D$ using $\beta_l^*=\beta_r\,$:
\bea
&\beta_r\bor\beta_l(\lambda)\,=\,S^{-1}(\lambda )
\qquad&\bor\beta_r\beta_l(h)=S(h)\label{BBrl}\\
&\beta_r\bor\beta_r(\lambda)\,=\,S(a\lac\lambda )
\qquad& \bor\beta_l\beta_l(h)=S^{-1}(h)\cdot a \label{BBrr}\\
&\beta_l\bor\beta_l(\lambda)\,=\,S(\alpha^{-1}\lambda )
\qquad&\bor\beta_r\beta_r(h)=S^{-1}(\alpha\lcoa h)\label{BBll}\\
&\beta_l\bor\beta_r(\lambda)\,=\,\alpha \cdot S^{-1}(\lambda\rac a )
\qquad&\bor\beta_l\beta_r(h)=a \cdot S(h\rcoa \alpha)\label{BBlr}
\eea
{}From (\ref{Sint})-(\ref{BBrl}) we can derive further useful relations
between adjoints:
\bea
\beta_l(S(h))\,=\,\omega\,a^{-1}\lac S(\beta_r(h))\qquad {\rm and} \qquad
&\beta_l(S^{-1}(h))\,=\,a^{-1}\lac S^{-1}(\beta_r(h))\label{SbetS}\\
\beta_l(\alpha\lcoa h)=\beta_r(S^2(h))\qquad {\rm resp} \qquad
&\mu((\alpha\lcoa k)h)=\mu(h S^2(k))\label{S2bet}\\
\bor\beta_l(S(\lambda))\,=\,\alpha^{-1}\lcoa S(\bor\beta_r(\lambda))\qquad {\rm
and} \qquad
&\bor\beta_l(S^{-1}(\lambda))\,=\,\omega \,\alpha^{-1}\lcoa
S^{-1}(\lambda)\label{SborbetS}\;.\\
\bor\beta_r(S^{-2}(\lambda))=\bor\beta_l(a\lac \lambda) \qquad {\rm resp}
\qquad & S^2\cot 1 \Delta'(x)=\Delta(x)(a\cot 1)\label{S2borbet}
\eea
Combining these identities we find Radfords formula for $S^4\,$:
\beq\label{S4}
S^4=ad^*(\alpha)\circ ad(a^{-1})
\eeq
Since in a double $\dA$ the adjoint action of $G(\oA)$ coincides with
the coadjoint action of $Ch(\A)$ on $\A$ we find from that (\ref{S4}) and
the corresponding equation on $\oA$ that the group like element $\alpha\cot
a^{-1}$ implements $S^4$ on $\dA\,$. The same is true  for the element $g$
defined in (\ref{deff}). In fact we have the following result of Drinfeld
for doubles:
\bpp\label{DrRd}{\rm [Dr1]}
$$g= \alpha\bdot a^{-1}$$
\epp

\bigskip
\ub{3.) Integrals and Canonical Traces of Doubles :}
In the construction of representations of the modular group  the integrals
 of the defining algebra play an important role.
The integral $\mu_D$ and the cointegral $x_D$ of a double $\dA\,$ clearly have
to be
 related to the integrals and cointegrals of $\A\,$. In this section we are
also interested in finding the moduli $\alpha_D$ and $a_D\,$. Comparing
Proposition \ref{DrRd} to
(\ref{S4}) we are led to expect that $\alpha_D=1$  and $a_D$ is the
same as $g\,$. We shall prove triviality of the comodulus first:
\bpp\label{pint}
For left integrals $\mu$ and $x$ as in (\ref{lints}) define the
canonical element in $\dA$  by $\;p\:=\mu\bdot S^{-1}(x)\,$.
Then
\ben
\item\qquad $S(p)=p$
\item $p$ is both a right and left cointegral in $\dA$ and $\alpha_D=1$
\item Let $P\in End(\A)$ be the image of $p$ under the map
$\bdot^{-1}:\dA\isto \A\cot\A^*=End(\A)\,$ is the projector onto the space of
left cointegrals and is given by:
\bea
\A
\stackrel{R\cot 1}{\hbox to 20pt{\rightarrowfill}}
\A\cot\oA \cot\A
\stackrel{1\cot \tau}{\hbox to 20pt{\rightarrowfill}}
\A\cot\A \cot\oA
\stackrel{\cdot \cot 1}{\hbox to 20pt{\rightarrowfill}}
\A \cot\oA
\stackrel{\Delta \cot 1}{\hbox to 20pt{\rightarrowfill}}
&\A\cot\A \cot\oA\nonumber\\
\stackrel{1\cot \tau}{\hbox to 20pt{\rightarrowfill}}
\A\cot\oA \cot\A
\stackrel{1\cot S^2}{\hbox to 20pt{\rightarrowfill}}
\A\cot\oA \cot\A
\stackrel{1\cot <,>}{\hbox to 20pt{\rightarrowfill}}
\A&\nonumber
\eea
\een
\epp
{\em Proof :} Since $S(p)=x\bdot S(\mu)$ and $S^{-1}(x)$ and $S(\mu)$ are right
integrals it is clear that {\em 1.)} implies that $p$ is an invariant with
respect to right and left multiplication of $\A$ and $\oA\,$. This shows {\em
2.)}. Assuming that {\em 1.)} is true we show {\em 3.)}:
\bea
\bowtie^{-1}\bigl(\mu\cot S^{-1}(x)\bigr)&=&
\Bigl<S(\mu'),S^{-1}(x)'\Bigr>\Bigl<\mu'',S^{-1}(x)'''\Bigr>\,S^{-1}(x)''\cot\mu''\label{bicp}\\
^{by (\ref{Sint})}&=&\omega^{-1}\Bigl<S(\mu'),S(x''')\Bigr>
\Bigl<\mu''',S(x')\Bigr>S(x'')\cot \mu''\nonumber\\
&=&\omega^{-1}\Bigl<\mu',x''\Bigr>\Bigl<\mu''',S(x')''\Bigr>S(x')'\cot
\mu''\nonumber\\
&=&\omega^{-1}\sum_{ij}\Bigl<\mu',x''\Bigr>\Bigl<\mu''',e_j''\Bigr>\Bigl<\mu'',e_i\Bigr>\Bigl<f_j,S(x')\Bigr>e_j'\cot f_i\nonumber\\
&=&\omega^{-1}\sum_{ij}\Bigl<\mu'',
e_j''e_i\Bigr>\Bigl<f_j,S\bor\beta_l(\mu')\Bigr>\,e_j'\cot f_i\nonumber\\
&=&\omega^{-1}\sum_{ij}\Bigl<f_j,S\bor\beta_l\beta_r\bigl(e''_je_i\bigr)\Bigr>\,e'_j\cot f_i\nonumber\\
^{by (\ref{BBrl})} &=&\omega^{-1}\sum_{ij}\Bigl<f_j,S^2(e''_je_i)\Bigr>\,
e'_j\cot f_i \qquad^{by({\em 1.)})}=x\cot S(m)\label{Pconv}
\eea

If we apply $1\cot S^2$ to both sides of the last equation and use (\ref{Sint})
we find
$$
P=\sum_{ij}\Bigl<f_j,S^2(e''_j)e_i\Bigr>e'_j\cot
f_i\,=\,\sum_j\Bigr<f''_j,S^2(e_j'')\Bigl>e_j'\cot
f_j'\quad^{by(\ref{dual2})}=\sum_{ij}\Bigr<f_j,S^2\bigl(e_j''e_i''\bigr)\Bigr>\,e_j'e_i'\cot f_i\;,
$$
which is precisely the equation given in {\em 3.)}. The same formula has been
proven in [Dr1] using the theory of Hopf modules directly.
{}From (\ref{Sint}) we find $S(\mu)(x)=\mu(ax)=1$ so that $P$ as in
(\ref{Pconv}) is a rank one projection.
It remains to show the first part of the proposition. For this purpose we need
two identities for the integrals, namely
\bea
S^2\Bigl(\bigl(S(\mu)''\rac a^{-1}\bigl)\alpha\Bigr)\cot
S(\mu)'&=&S^2\Bigl(\bigl(S(\mu)\rac a^{-1}\bigl)''\alpha\Bigl)\cot
\bigr(S(\mu)\rac a^{-1}\bigr)'\nonumber\\
^{by(\ref{Sint})}&=&S^2\bigl(\mu''\alpha\bigr)\cot \mu'\nonumber\\
^{by(\ref{S2bet})}&=&\mu' \cot \mu'' \qquad = \Delta(\mu)
\label{helpm}\eea
and
\bea
\Bigl(\bigl(S^2(x'')a^{-1}\bigr)\rcoa\alpha\Bigr) \cot x'
\quad^{by(\ref{S2borbet})}&=& \bigl((x')\rcoa\alpha\bigr)\cot x''\nonumber\\
&=&(x\rcoa\alpha)'\cot (x\rcoa \alpha)''\nonumber\\
^{by(\ref{Sint})}&=&S^{-1}(x)'\cot S^{-1}(x)''\quad
=\Delta\bigl(S(x)^{-1}\bigr)
\label{helpx}\eea
Inserting  (\ref{helpm}) and (\ref{helpx}) into the expression for the
bicrossed
product of $p$ in (\ref{bicp}) we find:
\bea
\bowtie^{-1}\bigl(\mu\cot S^{-1}(x)\bigr) &=&\Bigr<S(\mu') ,
S^{-1}(x)'\Bigr>\:\Bigl<\mu''',S^{-1}(x)'''\Bigr>\:S^{-1}(x)''\cot
\mu''\nonumber\\
&=&\Bigr<S^3\Bigl(\bigl(S(\mu)\bigr)'''\rac
a^{-1}\bigr)\alpha\Bigr),\,S^2\bigl((x''a^{-1})\rcoa\alpha\bigr)\Bigr>\:
\Bigl<S(\mu)'',x''\Bigr>\:x'\cot S(\mu)'\nonumber\\
&=&\Bigl<\alpha^{-1}\Bigl(a\lac
S\bigl(S(\mu)'''\bigr)\Bigr),\,\bigl(x'''a^{-1}\bigr)\rcoa\alpha\Bigr>\:\Bigl<S(\mu)'',x''\Bigr>\:x'\cot S(\mu)'\nonumber\\
&=&\Bigl<S\bigl(S(\mu)'''\bigr),\,x'''\Bigr>\:\Bigl<S(\mu)'',x''\Bigr>\:x'\cot
S(\mu)'\nonumber\\
&=&\Bigl<S(\mu)''S\bigl(S(\mu)'''\bigr),\,x''\Bigr>\:x'\cot
S(\mu)'\qquad\quad=\;x\cot S(\mu)\;.\label{pres}
\eea
Hence $\mu\bdot S^{-1}(x)=S\bigl(\mu\bdot S^{-1}(x)\bigr)=x\bdot S(\mu)$ and we
have shown {\em 1.)} of the proposition. \hfill$\Box$

\medskip

Thanks to the simple comultiplicative structure of $\dA$ it is much easier
 to find the integral. Since $S^{-1}(\mu)$ is a right integral of $\A$ and
since
we have opposite comultiplication on $\oA$ a {\em right} integral of $\dA$ is
given by
\beq\label{defDint}
\mu_D\bigl(\lambda \bdot y\bigr)\,:=\,\lambda(x)\mu\bigl(S(y)\bigr)\qquad
\hbox{for all}\quad \lambda\in\oA,\,\hbox{ and }\, y\in\A\;.
\eeq
 Its properties are described next:
\bpp\label{chDint}
Let $x$ and $\mu$ be the left integrals of $\A$ as in (\ref{lints}). Then
\ben
\item $\mu_D$ is a right integral of $\dA$ with $\mu_D(p)=1\,$.
\item The modulus (as defined in \ref{modrel}) of $\dA$ is $a_D=g^{-1}\,$.
\item $\mu_D\in\bor C\bigl(\dA\bigr)\,$, in particular $\mu_D(\lambda\bdot
y)=\omega \mu_D(y\bdot\lambda)\;\forall\, y\in\A,\, \lambda\in \oA\;$.
\een
\epp
{\em Proof :} Part {\em 1.)} is clear. Also, it follows directly from the
definitions of the moduli of $\A$ that $1\cot\mu_D\,\Delta(h)=\alpha\bdot
a^{-1}\mu_D(h)\,$. {\em 2.)} follows if we apply the antipode and use that
$S(\mu_D)$ is a left integral of $\dA\,$. In order to show {\em 3.)} we
observe that the equation for right integrals analogous to (\ref{S2bet})
is $\mu_D\bigl(S^2(k)h\bigr)=\mu_D\bigl(h(k\rcoa\alpha_D^{-1})\bigr)\,$.
Together with Proposition \ref{pint} {2.)} this shows $\mu_D\in\bor
C\bigl(\dA\bigr)\,$. In the case where $k$ and $h$ are in
the special subalgebras we use (\ref{Sint}) to show
$\mu_D(S^2(\lambda)y)=\omega^{-1}\mu_D(\lambda y)\,$ which yields the last
equation in part {\em 3.)}.

\hfill $\Box$

 The fact that the right integral of a double algebra is invariant under the
coadjoint action allows us to identify as an object in the representation
category of $\dA\,$, namely with the integral of the``braided algebra'' [Ly] of
the category. Before we explain this aspect in more detail in the next section
let us discuss a few more consequences of Proposition \ref{chDint} for doubles.

It is easy to see that an element of a Hopf algebra $w\in D$ with
$S^2(y)=wyw^{-1}$ provides us with an isomorphism $\bor C(D) \isto
C_o(D)\,:\,\lambda\mapsto\lambda\rac w\,$. Here $C_o(D)$ denotes as in [Dr1]
the space of traces on $D\,$. Given the two canonical elements in (\ref{relu})
we
wish to compute the respective traces for $\mu_D$. To this end define the
following linear forms on $\dA\,$:
\bea\label{chi}
\chi\,&:&\,\dA\stackrel{\bdot^{-1}}{\hbox to 25pt{\rightarrowfill}}\oA\cot\A
\stackrel{<,>}{\hbox to 15pt{\rightarrowfill}}k\nonumber\\
\hat\chi\,&:&\,\dA\stackrel{\bdot^{-1}}{\hbox to
25pt{\rightarrowfill}}\oA\cot\A
\stackrel{1\cot S^{-1}}{\hbox to 30pt{\rightarrowfill}}
\oA\cot\A
\stackrel{<,>}{\hbox to 25pt{\rightarrowfill}}k
\eea
The forms on $\dA$ and the canonical elements are now related as follows.
\bpp\label{cantr}
\beq
\chi\,=\,\mu_D\rac u \; ,\qquad\quad \hat\chi\,=\,\omega^{-1}\,\mu_D\rac\hat
u\;,
\eeq
and both $\chi$ and $\hat\chi$ are \underline{nondegenerate traces} on $\dA\,$.
\epp
{\em Proof :} From previous considerations it is clear that $\mu_D\rac u$
and $\mu_D\rac\hat u$ are traces. The rest of the proof are straightforward
computations:
\bea
(\mu_D\rac u)\bigl(\lambda\bdot y\bigr)&=&(\mu_D \rac u)\bigl(y\bdot
\lambda\bigr)
\quad=\sum_i\mu_D\bigl(S(f_i)\bdot e_i\cdot y\bdot \lambda\bigr)\nonumber\\
&=&\sum_i\mu_D\bigl(S^2(\lambda)\cdot S(f_i)\bdot e_i\cdot y\bigr)
\quad=\sum_i
\bigl(S^2(\lambda)S(f_i)\bigr)(x)\:\mu\bigl(S(y)S(e_i)\bigr)\nonumber\\
&=&\,\sum_i\,f_i\bigl(x\rcoa S^2(\lambda)\bigr)\:\mu\bigl(S(y)e_i\bigr)
\quad=\,\mu\Bigl(S(y)\bigl(x\rcoa S^2(\lambda)\bigl)\Bigl)\nonumber\\
&=&\mu\Bigl(S(y)\bor\beta_l\bigl(S^2(\lambda)\bigr)\Bigr)
\quad=\beta_r\bor\beta_l\bigl(S^2(\lambda)\bigr)\bigl(S(y)\bigr)\nonumber\\
^{by (\ref{BBrl})}
&=&S(\lambda)\bigl(S(y)\bigr)
\quad =\,\lambda(y)\quad =\,\chi(\lambda\bdot y )\qquad.\nonumber
\eea
Similarly,
\bea
(\mu_D\rac\hat u)(\lambda\bdot y)&=& \sum_i \mu_D\bigl(f_i\cdot\lambda\bdot
y\cdot e_i\bigr) \quad= \sum_i f_i\cdot
\lambda(x)\:\mu\bigl(S(ye_i)\bigr)\nonumber\\
&=&\mu\Bigl(S\bigl(y(\lambda\lcoa x)\bigr)\Bigr)\quad ^{by
(\ref{Sint})}=\,\omega\,\mu\bigl(ay(\lambda\lcoa x)\bigr)\nonumber\\
&=&\omega\,\mu\bigl(ay\bor\beta_r(\lambda)\bigr)\quad = \;\omega
\,\beta_r\bor\beta_r(\lambda)(ay)\nonumber\\
^{by(\ref{BBrr})}&=&\;\omega\, S(a\lac \lambda)(ay)\quad = \; \omega\,
S(\lambda)(y) \quad = \; \omega\,\hat\chi(\lambda\bdot y)\qquad.\nonumber
\eea
Nondegeneracy of $\chi$ and $\hat\chi$ follow directly from nondegeneracy of
$\mu_D\,$.

\hfill$\Box$

\bigskip
\ub{4.) Balancing in Doubles :}
In a rigid BTC any object $X$ is isomorphic to its double conjugate
$X^{\vee\vee}$. Yet the only isomorphism that is a priori  canonical
is between $X$ and  $X^{\vee\vee\vee\vee}\,$. Thus in addition to the usual
axioms defining a BTC one often requires the existence of a
$\otimes -$ natural isomorphism of the functor $X\to  X^{\vee\vee}$ to the
identity, which squares to the canonical one from $X\to X^{\vee\vee\vee\vee}$
to the identity. For the representation category of a quasitriangular Hopf
algebra $D$ this is
equivalent to the existence of a group like element $k$ with:
\beq\label{relk}
k\in G(D)\,, \qquad g=k^2\,, \quad {\rm and} \quad S^2(y)=kyk^{-1}\;.
\eeq
It is clear that a  balancing does not have exist since often $g$ is not a
square in $G(D)\,$.
If it does it is unique up to multiplication with central, group like elements
of order two, i.e., elements in $\Sigma(D):=\,_2\!G(D)\cap Z(D)\,$.

Equivalently, we can consider the corresponding element $v:=u\cdot k^{-1}=\hat
u^{-1}\cdot k\,$. Inspecting (\ref{defz}) it is easily verified that $v$
defines a balancing elements iff
\beq\label{ribbon}
v\in Z(D)\,, \qquad S(v)=v\,, \quad \hbox{and}\quad M = v\cot v \Delta(v^{-1})
\eeq
{}From these conditions  $\epsilon(v)=1$ and $v^2=z\,$ follow.
This point of view has been introduced in [RT0] where $v$ is called a  {\em
ribbon element}. In their context the eigenvalue of $v$ in an irreducible
representation yields the framing anomalies of
colored link.

For a double $\dA$ the existence of a balancing can be phrased as a property
of the moduli of $\A\,$.

\bpp\label{moduli}
\hfill
\ben
\item $k$ is a balancing of $\dA$ if and only if
\bea
k&=&\sqrt\alpha\bdot (\sqrt a)^{-1}\label{kform}\\
&&\hbox{where}\quad \sqrt\alpha\in G(\oA)\,,\;\sqrt a \in G(\A)\quad
\hbox{square to}\;\alpha\hbox{ and } a \;\hbox{ respectively}\nonumber\\
\hbox{and}&&\nonumber\\
S^2\,&=&\,ad^*(\sqrt\alpha)\circ ad(\sqrt a^{-1})\qquad\hbox{  on }
\quad\A\qquad\label{S2k} .
\eea
\item To a given balancing we associate the  number $\nu$ defined by
\beq\label{nu}
\nu^{-1}\,=\,\chi(k)\,=\,\sqrt\alpha(\sqrt a )\;.
\eeq
If $\dA$ admits a balancing $\nu$ is a root of unity, $\nu^4=\omega$ and
\newline\quad $\nu^2$ does not depend on the choice of balancing.
\een
\epp
{\em Proof :} From Lemma \ref{dougrou} and (\ref{relk}) we infer that $k$
has to be a product of group likes of the special subalgebras. By definition
of doubles (\ref{S2k}) is the same as $S^2(y)=kyk^{-1}\;\forall\,y\in\A\,$.
The inverse adjoint of (\ref{S2k}) yields the same equation on $\oA$ and
thereby (\ref{relk}). For part {\em 2.)} we remark that two balancings $k$ and
$k'$ are related by $k'=k\bdot R$ where $R=\rho\bdot
r\,\in\,\Sigma(\dA)\cong\Sigma(\oA)\oplus\Sigma(\A)\,$. Then
$\chi(k')\cdot\chi(k)^{-1}\,=\,\sqrt\alpha(r)\,\rho(\sqrt a)\,\rho(r)\;$ which
is of order two since $\rho$ and $r$ are.
\hfill$\Box$

In particular, the last statement implies that once a balancing exists the
intrinsic quantity $\omega$  has a canonical square root.

\bigskip

\ub{5.) Representation of Mapping Class Groups on Doubles :}

In several papers [Ly] Lyubachenko has developed the notion of a Hopf algebra
$F$
in a braided tensor category $\cal C$. It is analogue to the notion of braided
group, as defined by Majid [M2].  As an
object $F$ in a category with all limits the algebra is  the constant functor
of the coend \mbox{$\Bigl<\inth\,;\,h:\inth\stackrel {\hbox{ \Large ..  }
}{\hbox to 16pt{\rightarrowfill } } F \Bigr>$ } of the functor $\inth\,:\,{\cal
C }^{opp}\times{\cal C}\to {\cal C}\,:\,(X,Y)\mapsto X^{\vee}\cot Y\,$.
For definitions see [Mc].
The multiplication and comultiplication of $F$ are induced by certain
compositions of dinatural transformations using universality of the coend.
As opposed to symmetric categories the definition of the multiplication of $F$
depends on the choice of a commutativity isomorphism. The same is true for the
axiom replacing cohomomorphie of the multiplication. An analogous statement of
the fundamental theorem of Hopf modules holds for the braided algebras so that
under certain finiteness conditions the algebra has an integral $\mu\in
Hom(1,F)\,$. The algebra also possesses a braided antipode  $\Gamma \in
End(F)\,$. Lyubachenko constructs, in analogy to the definitions for semisimple
categories, modular operators ${\cal T},\,{\cal S}\,\in End(F)\,$. They are
determined by the coend properties of $F$ and the following commutative
diagrams:
\beq\bar{ccccccc}\label{TandG}
X^{\vee}\cot X
&
\stackrel{1\cot v_X  }{\hbox to 25pt{\rightarrowfill}}
&
X^{\vee}\cot X
&
\hphantom{xxxxxxxxxxx}
&
X^{\vee}\cot X
&
\stackrel{\gamma_X}{\hbox to 25pt{\rightarrowfill}}
&
X^{\vee\vee}\cot X^{\vee}
\\
\quad\Bigg\downarrow h_X
&&
\quad\Bigg\downarrow h_X
&&
\quad\Bigg\downarrow h_X
&&
\quad\Bigg\downarrow h_{X^{\vee}}
\\
F
&
\stackrel{\cal T}{\hbox to 25pt{\rightarrowfill}}
&
F
&
&
F
&
\stackrel{\Gamma}{\hbox to 25pt{\rightarrowfill}}
&
F
\ear
\eeq
Here $v_{-}\in Nat(id)$ is the balancing and $\gamma_X:=\,q_X\cot 1\,
\epsilon(X^{\vee},X)\,$, where $\epsilon$ is the commutativity constraint and
$q_X\,:\,X \to X^{\vee}\cot X^{\vee\vee}\cot X
\stackrel{1\cot \epsilon}{\hbox to 20pt{\rightarrowfill}}
X^{\vee}\cot X \cot X^{\vee\vee} \to X\,$. Furthermore,
\beq\bar{ccccl}\label{comS}
Y^{\vee}\cot Y
&
\stackrel{\mu \cot 1}{\hbox to 20pt{\rightarrowfill}}
&
F\cot Y^{\vee}\cot Y
&
\stackrel{h_X \cot 1}{\hbox to 20pt{\leftarrowfill}}
&
X^{\vee}\cot X\cot Y^{\vee}\cot Y \\
\Bigg\downarrow
&&
\Bigg\downarrow
&&
\qquad\quad\Biggr\downarrow(1\!\cot\!1\!\cot\!
ev)(1\!\cot\!\epsilon^2\!\cot\!1)
\\
F
&
\stackrel{\cal S}{\hbox to 30pt{\rightarrowfill}}
&
F
&
\stackrel{h_X}{\hbox to 30pt{\leftarrowfill}}
&
\quad X^{\vee}\cot X\qquad .
\ear
\eeq
The coend and integral exist if ${\cal C}$ is the representation category of a
finite dimensional  Hopf algebra $D\,$. Specifically, we have that $F=D^*\,$,
which is a $D-$module by $ad_*$-action. The comultiplication is just the
multiplication on $D$. However, the multiplication in $F$ stems from a
distorted
coproduct $\Delta_B$ on $D$ as the usual one is not $ad_*$-covariant.
In one convention we have, e.g., $\Delta_{Br}(y)\,=\, e_i''y'S^{-1}(e_i')\cot
f_iy''\,$.  As remarked in [LyM] the right integral for the braided
multiplication coincides with the ordinary right integral. This is seen easily,
e.g., from the fact that $\mu_D$ for a double is $ad_*$-invariant.
An antipode $\Gamma\in End(D)\,$ of the braided multiplication consistent with
$\Delta_B$ is
\beq\label{BrS}
\Gamma (A)\,:=\,\sum_i \,S(e_i)S(A)\hat u f_i\;.
\eeq
The triple $\Bigl( D,\cdot, \Delta_{Br}\Bigr)$ is the prototype of a braided
group. For an thorough teatment of this structure which inspired
the algebra construction in [Ly] we refer to [M2]. However, the construction of
the $\cal S$ and $\cal T$ given in [Ly] can also be translated into the context
of ordinary Hopf algebras. The action of $\cal T$ is clearly given by
multiplication of a ribbon element $v\,$. The identity of integrals allows us
to derive
from (\ref{comS})  a formula for $\cal S$ acting on
a quasitriangular algebra $D\,$.
\beq\label{ffcS}
{\cal S}(A)\,=\,S\bigl(\bor f(\mu_D\rac
A)\bigr)\;=\,\sum_{i,j}\mu_D\bigl(Af_je_i\bigr)S(f_i)S(e_j)
\eeq
This formula (with slightly different conventions) has been given in [LyM].
Using the
form of the right integral given in (\ref{chDint}) and applying the bicross
formula
(\ref{bic+}) to order $Af_je_i$ this formula can be worked out further. The
formula ${\cal S}(\lambda\cot h)\,=\,\sum_i f_i\cot
(x'')<\lambda,e_i'''x'S^{-1}(e'_i)>$ resulting
from this has been given in [M1].

\medskip
Let us now use the properties of integrals given in the previous section and
the identities for the canonical isomorphisms to derive an intriguingly,
compact formula  for $\cal S\,$.
{}From this form the invertibility of $\cal S$ for doubles is obvious and the
inverse readily
computed from the identities (\ref{BBrr}) and following.

\bpp\label{Smatrix}
For a double $\dA$ over a finite dimensional Hopf algebra $\A$ let $\mu_D$ be
as in (\ref{defDint}) and ${\cal S}\,\in\,End(\dA )$ be defined as in
(\ref{ffcS}). Then
the following diagrams of isomorphisms commute:
\beq\bar{ccccccc}\A\cot \oA
&
\raise 1.37ex\hbox{$\underline{ \beta_l\circ S^{-1}\,\cot
\,S\circ\bor\beta_l}$}\mkern -6mu\to
&
\oA\cot\A
&
\hphantom{xx}
&
\oA\cot\A
&
\raise 1.37ex\hbox{$\underline{ \bor\beta_r\,\cot\,\beta_l\circ L_a}$}\mkern
-6mu\to
&
\A\cot \oA
\\
\Bigg\downarrow \bdot
&&
\Bigg\downarrow \bdot
&&
\Bigg\downarrow \bdot
&&
\Bigg\downarrow \bdot
\\
\dA
&
\raise .83ex\hbox{$\underline{ \,\,\quad\qquad { \cal S  }
\,\,\quad\qquad}$}\mkern -6mu\to
&
\dA
&
\hphantom{xx}
&
\dA
&
\raise .83ex\hbox{$\underline{ \,\,\qquad { {\cal S}^{-1}  }
\,\,\qquad}$}\mkern -6mu\to
&
\dA
\\
&&&&&&\label{smat}
\ear
\eeq
Here $L_a$ denotes left multiplication with $a\,$.
\epp
{\em Proof :} The first diagram is verified by direct computation:
\bea\label{}
{\cal S}(y\bdot \lambda)\,&=&\,\sum_{ij}\mu_D\bigl(y\bdot\lambda
f_je_i\bigr)\cot S(f_i)S(e_j)\;\quad\quad ^{by(Pp
\ref{chDint}.3.)}=\,\sum_{ij}\mu_D\bigl(\lambda f_j\bdot
e_iS^{-2}(y)\bigr)\nonumber\\
^{by(\ref{defDint})}&=& \sum_{ij}\lambda
f_j(x)\,\mu\bigl(S^{-1}(y)e_i\bigr)\,f_i\bdot
S(e_j)\,\qquad\qquad\qquad=\mu\rac S^{-1}(y)\bdot S(x\rcoa\lambda)\nonumber\\
^{by(\ref{defbet})}&=&\beta_l(S^{-1}(y))\bdot S\bigl(\bor\beta_l(\lambda)\bigr)
\eea
The second diagram follows immediately from relations (\ref{BBrl}) and
(\ref{BBrr}), which allow us to invert $\beta_l$ and $\bor\beta_l\,$.
\hfill$\Box$

\medskip
Let us also give a more convenient form for the braided antipode:
\blm\label{Brant}
Let  $\tilde\Gamma \,:\,\oA\cot\A\isto\A\cot\oA$ be given by
\bea
\oA\cot\A
\stackrel{R\cot 1^2\cot R}{\hbox to 35pt{\rightarrowfill}}
\A\cot\oA\cot\oA\cot\A\cot\A\cot\oA
\stackrel{S\cot 1\cot S^{-1} \cot 1^3  }{\hbox to 35pt{\rightarrowfill}}
\A\cot\oA\cot\oA\cot\A\cot\A\cot\oA\nonumber\\
\stackrel{1\cot \cdot \cot \cdot \cot 1}{\hbox to 35pt{\rightarrowfill}}
\A\cot\oA\cot\A\cot\oA
\stackrel{1\cot \tau \cot 1  }{\hbox to 30pt{\rightarrowfill}}
\A\cot\A\cot\oA\cot\oA
\stackrel{\cdot \cot \cdot  }{\hbox to 20pt{\rightarrowfill}}
\A\cot\oA
\stackrel{ S \cot 1 }{\hbox to 25pt{\rightarrowfill}}
\A\cot\oA\nonumber
\eea
and $\Gamma$ as in (\ref{BrS}). Then the following diagram commutes:
\beq
\bar{ccc}
\oA\cot\A
&
\raise .83ex\hbox{$\underline{ \,\,\qquad { \tilde \Gamma  }
\,\,\qquad}$}\mkern -6mu\to
&
\A\cot\oA
\\
\Bigg\downarrow \bdot
&&
\Bigg\downarrow \bdot
\\
\dA
&
\raise .83ex\hbox{$\underline{ \,\,\qquad { \Gamma  } \,\,\qquad}$}\mkern
-6mu\to
&
\dA
\ear
\eeq
\elm
{\em Proof :} Straightforward computation:
\bea
\Gamma(\lambda\bdot y) &=& \sum_i\,S(e_i)S(y)\bdot S(\lambda)\hat u f_i\qquad
=\,\sum_iS(e_i)S(y)\hat u S^{-1}(\lambda)f_i\nonumber\\
&=&\,\sum_{ij} S\bigl(S(f_j)ye_i\bigr)\bdot
e_jS^{-1}(\lambda)f_i\;,\label{BrSexp}
\eea
which is precisely the above composition.
\hfill$\Box$

\bigskip
\ub{6.) Proof of Modular Relations and the Projective Phases :}

For the square of the braided antipode we easily verify
\beq\label{gam2}
\Gamma^2\,=\,ad^-(v^{-1})\;,
\eeq
where $v$ is any ribbon element and $ad^-(y)=S^{-1}\circ ad(y)\circ S\,$.
Proposition \ref{Smatrix} and Lemma \ref{Brant} put us in a position to prove
the
next lemma. From this we will infer one of the modular relations and the
correct projective phase.
\blm
We have the following relation for maps $\oA\cot\A\,\isto\,\oA\cot\A\,$:
\beq\label{morelex}
\beta_l\!\circ\!S^{-1}\! \cot\! S\!\circ\!\bor\beta_l \;\tilde\Gamma\quad
=\quad \omega\;\bowtie\;\bor\beta_r\!\cot\! (\beta_lL_a)
\eeq
\elm

{\em Proof :} We shall prove (\ref{morelex}) by evaluating both sides on
$\lambda\cot y\in \oA\cot\A $ individually and comparing results. For the right
hand side we have

\bea
\bowtie\,\bor\beta_r\!\cot\!(\beta_lL_a)(\lambda\cot
y)\;&=&\;\bowtie\,\Bigl(x'\lambda(x'')\cot \mu\rac(ay)\Bigr) \nonumber\\
&=&\Bigl<\bigl(\mu\rac(ay)\bigr)',x'\Bigr>\,\Bigl<S\Bigl(\bigl(\mu\rac(ay)\bigr)'''\Bigr),x'''\Bigr>\,
\lambda(x'''')\;\bigl(\mu\rac (ay)\bigr)''\cot x''\nonumber\\
&=&\Bigl<\mu',x'\Bigr>\,\Bigl<\mu''',ayS^{-1}(x''')\Bigr>\,\lambda(x'''')\:
\mu''\cot x''\nonumber\\
&=&\,\sum_i\,\Bigl<\mu',x'\Bigr>\,\Bigl<f_i,x''\Bigr>
\,\Bigl<\mu''',ayS^{-1}(e_i'')\Bigr>\,\lambda(e_i''')\: \mu''\cot
e_i'\nonumber\\
&=&\,\sum_i\,\Bigl<\mu',\bor\beta_r(f_i)\Bigr>\,\Bigl<\mu''',ayS^{-1}(e_i'')\Bigr>
\,\lambda(e_i''')\:\mu''\cot e_i'\nonumber\\
&=&\sum_i\Bigl<\beta_r\bor\beta_r(f_i)'',\,ayS^{-1}(e_i'')\Bigr>\lambda(e_i''')\,
\beta_r\bor\beta_r(f_i)' \cot e_i' \nonumber\\
^{by(\ref{BBrr})}&=&\sum_i\,\Bigl< \bigl(S(f_i)\rac
a^{-1}\bigr)'',ayS^{-1}(e_i'')\Bigr>\,\lambda(e_i''')\: \Bigl(S(f_i)\rac a^{-1}
\Bigr)' \cot e_i' \nonumber\\
&=&\sum_i\Bigl<S( f_i)'',yS^{-1}(e_i'')\Bigr>\,\lambda(e_i''')\: S(f_i)'\cot
e_i'\qquad\quad
\label{ret1}\eea
\noindent
The evaluation of the left hand side gives:
\bea
\beta_l\!\circ\!S^{-1}\! \cot\! S\!\circ\!\bor\beta_l&
\:\tilde\Gamma&\bigl(\lambda\cot
y\bigr)\quad=\quad\sum_{ij}\,\beta_l\!\circ\!S^{-1}\! \cot\!
S\!\circ\!\bor\beta_l\,\Bigl(S\bigl(S(e_i)ye_j\bigr)\cot\bigr(f_iS^{-1}(\lambda)f_j\bigr)\Bigr)\nonumber\\
&=&\sum_{ij}\beta_l\bigl(S(e_i)ye_j\bigr)\cot
S\circ\bor\beta_l\bigl(f_iS^{-1}(\lambda)f_j\bigr)\nonumber\\
&=&\sum_{ij}\Bigl<f_i,x'\Bigr>\,\Bigl<S^{-1}(\lambda)f_j,x''\Bigr>\:\mu\rac\bigl(S(e_i)ye_j\bigr)\cot S(x''')\nonumber\\
&=&\sum_{j}\,\Bigl<S^{-1}(\lambda)f_j,x''\Bigr>\:S\Bigl(\bigl(S^{-1}(ye_j)x'\bigr)\lac S^{-1}(\mu)\Bigr)\cot S(x''')\nonumber\\
^{by(\ref{Sint})}
&=&\omega\,\sum_{ij}\,\Bigl<f_i,x''\Bigr>\,\Bigl<S^{-1}(\lambda)f_j,e_i'\Bigr>\:S\Bigl(\bigl(S^{-1}(ye_j)x'\bigr)\lac \mu \rac a\Bigr)\cot S(e_i'')\nonumber\\
&=&\omega\,\sum_{ij}\,\,\Bigl<S^{-1}(\lambda)f_j,e_i'\Bigr>\:S\Bigl(\bigl(S^{-1}(ye_j)\bor\beta_r(f_i)\bigr)\lac \mu \rac a\Bigr)\cot S(e_i'')\nonumber\\
&=&
\omega\,\sum_{ij}\,\,\Bigl<S^{-1}(\lambda)f_j,e_i'\Bigr>\:S\Bigl(\bigl(S^{-1}(ye_j)\bigr)\lac \bigl(\beta_r\bor\beta_r(f_i)\bigr)\rac a\Bigr)\cot S(e_i'')\nonumber\\
^{by(\ref{BBrr})}
&=&
\omega\,\sum_{ij}\,\,\Bigl<S^{-1}(\lambda)f_j,e_i'\Bigr>\:S\Bigl(\bigl(S^{-1}(ye_j)\bigr)\lac S\bigl(a\lac f_i\bigr)\rac a\Bigr)\cot S(e_i'')\nonumber\\
&=&
\omega\,\sum_{ij}\,\Bigl<S^{-1}(\lambda)f_j,e_i'\Bigr>\:S^2(f_i)\rac (ye_j)\cot
S(e_i'')\nonumber\\
&=&
\omega\,\sum_{ij}\,\Bigl<S(f_j)\lambda ,e_i''\Bigr>\:S(f_i)\rac (ye_j)\cot
e_i'\nonumber\\
&=&
\omega\,\sum_{ij}\,\Bigl<S(f_i)'',ye_j\Bigr>\,
\Bigl<S(f_j),e_i''\Bigr>\,\lambda(e_i''')\:S(f_i)'\cot e_i'\nonumber\\
&=&
\omega\,\sum_{i}\,\Bigl<S(f_i)'',yS^{-1}(e_i'')\Bigr>\,
\lambda(e_i''')\:S(f_i)'\cot e_i' \label{ret2}
\eea
Comparison of (\ref{ret1}) to (\ref{ret2}) proves the assertion. \hfill$\Box$

The projective phases of the second modular relation arise in the computation
of the value of $\cal S$ on the ribbon element.

\blm\label{Sonv}
Suppose $v$ is a ribbon element of a double $\dA$ and $\nu$ is the associated
fourth root of $\omega$ (see Prop.\ref{moduli}.).  Then we have for $\cal S$
as defined in (\ref{ffcS}) :
\beq
\qquad{\cal S}(v)\,=\,\nu^{-1}\,v^{-1}   \qquad\qquad {\cal
S}(v^{-1})\,=\,\nu^5\,v\quad .
\eeq
\elm

{\em Proof :} Straightforward computation: Using $v=uk^{-1}$ we have
\bea\label{cccc1}
{\cal S}(v)\,&=&\,\sum_{ij}\,\mu_D\bigl(uk^{-1}f_je_i\bigr)\,S(f_i)\bdot
S(e_j)\qquad ^{by
Prop\ref{cantr}.}=\,\sum_{ij}\,\chi\bigl(e_ik^{-1}f_j\bigr)\,S(f_i)\bdot S(e_j)
\nonumber\\
&=&\,\sum_{ij}\,\Bigl<\sqrt\alpha^{-1}f_j, e_i\sqrt a\Bigr>\,S(f_i)\bdot S(e_j)
\qquad\qquad =\,\sum_{i}\, S(f_i)\bdot S\Bigl((e_i\sqrt a)\rcoa
\sqrt\alpha^{-1}\Bigr)\nonumber\\
&=&\nu^{-1}\,\sum_{i}\, S\bigl(f_i\bigr)\bdot\sqrt a^{-1}S\Bigl(e_i\rcoa
\sqrt\alpha^{-1}\Bigr)\quad\qquad=\nu^{-1}\, \sum_{i}\, S\bigl(\sqrt\alpha^{-1}
        f_i\bigr)\bdot\sqrt a^{-1}S(e_i)\nonumber\\
&=&\nu^{-1}\,\sum_{i}\,S(f_i)k S(e_i) \;=\,\nu^{-1}\,\sum_{i}\,f_i S^2(e_i)k\;
=\;\nu^{-1}\,u^{-1}k\;=\;\nu^{-1}\,v^{-1}\qquad.
\eea
The second relation follows with $v^{-1}=\hat u k^{-1}$ from:
\bea\label{cccc2}
\quad {\cal S}(v^{-1})&=&\sum_{ij}\,\mu_D\bigl(\hat u k^{-1} f_je_i\bigr)\,
S(f_i)\bdot S(e_j)\quad=\,\omega\, \sum_{ij}\,\hat\chi
\bigl(f_j\sqrt \alpha^{-1} \sqrt a S^2(e_i)\bigr)\, S(f_i)\bdot
S(e_j)\nonumber\\
&=&\,\omega\,\sum_{ij}\,\Bigl<f_j\sqrt\alpha^{-1}, S(e_i)\sqrt a ^{-1}\Bigr>\,
S(f_i)\bdot S(e_j)\;=\,\omega\sum_i\,f_i\bdot S\Bigl(\sqrt
\alpha^{-1}\lcoa\bigl(e_i\sqrt a ^{-1}\bigr)\Bigr)\nonumber\\
&=&\,\omega\nu\sum_i\,f_i\bdot\sqrt a S\bigl(\sqrt\alpha^{-1}\lcoa
e_i\bigr)\quad = \, \omega\nu\,\sum_i\,f_i\sqrt\alpha^{-1}\bdot \sqrt a
S(e_i)\;\nonumber\\
&=&\nu^5k^{-1}u\quad=\,\nu^5\,v \qquad .
\eea

Let us now prove the second modular relation:
\bpp\label{sts}
For a double $\dA$ with balancing, let $\cal S$ be defined as in (\ref{ffcS})
and $\cal T$ by multiplication with $v\,$. Then
\beq \label{ststst}
{\cal S}{\cal T}^{-1}{\cal S}\;=\;\nu^5\,{\cal T}{\cal S}{\cal T}
\eeq
\epp

{\em Proof :} If we apply $\eta\circ S^{-1}\cot 1 \cot S$ for some $\eta\in
\dA^*$ to both sides of the equation in Lemma \ref{Meq} we find with
$\tau\bigl(M\bigr)\,=\,S\cot S \bigl(M\bigr)\,$ that
\bea\label{calc1}
S\circ\bor f\Bigl(\lambda\rac S\circ\bor f(\eta)\Bigr)&=&\,\bigl(\lambda\cot
S\bigr)\Bigl(\bigl(S(\bor f(\eta))\cot 1 \bigr)\,M\Bigr)\nonumber\\
&=&\,\eta\circ S^{-1}\cot \lambda \cot S \,\Bigl(\bigl(\tau(M)\cot 1\bigr)
\bigl(1\cot M\bigr)\Bigr)\nonumber\\
&=&\,\sum_{ij}\eta\Bigl(S^{-1}\bigl(\bor f(\lambda)'f_if_j\bigr)\Bigr)\cot
S(e_i)S\bigl(\bor f(\lambda)''\bigr)\,e_j
\eea

Inserting into (\ref{calc1}) the forms $\,\lambda\,=\,\mu_D\rac\rho\,$ and
$\,\eta\,=\,\mu_D\rac A\,$ for some $A,\rho\in\dA$ and by using the definition
(\ref{ffcS}) we find:
\bea\label{calc2}
{\cal S}\circ L_{\rho}\circ {\cal S}(A)\,&=&\,\sum_{ij}\,
\mu_D\Bigl(A\,S^{-1}\Bigl(\Bigl(S^{-1}\bigl({\cal
S}(\rho)\bigr)\Bigr)'f_if_j\Bigr)\Bigr)
\,S(e_i)S\Bigl(S^{-1}\bigl({\cal S}(\rho)\bigr)''\Bigr)\,e_j\nonumber\\
&=&\sum_{ij}\,\mu_D\Bigl(A\,S^{-1}(f_j)f_iS^{-2}
\bigl({\cal S}(\rho)''\bigr)\Bigr)\,S^2(e_i){\cal S}(\rho)'e_j
\eea

Here $L_{\rho}\,$ is the left multiplication with $\rho\,$. The left hand side
of the assertion (\ref{ststst}) is now found by
 specializing $\rho=v^{-1}$ where $L_v=\cal T\,$. In order to evaluate the
right hand side of (\ref{calc2}) we notice that Lemma \ref{Sonv} implies the
following identities:
\bea
\Delta\bigl({\cal S}(v^{-1})\bigr)\,&=&\,\nu^5\Delta(v)\,=\,\nu^5 v\cot v
M^{-1}\,=\,\nu^5v\cot v\, R^{-1}\tau\bigl(R^{-1}\bigr)\nonumber\\
&=&\,\nu^5v\cot v \, S^3\cot S^2\bigl(R\bigr)\,1\cot
S\bigl(\tau(R)\bigr)\,=\,\nu^5\,\sum_{kl}vS^3(e_k)f_l\cot
vS^2(f_k)S(e_l)\nonumber
\eea
Replacing ${\cal S}(v^{-1})'\cot {\cal S}(v^{-1})''\,$ in (\ref{calc2}) by this
expression yields the assertion:
\bea\label{finally}
{\cal S}{\cal T}^{-1}{\cal S}\,
&=&\nu^5\,\sum_{ijkl}\mu_D\Bigl(A\,S^{-1}(f_j)f_iv f_k
S^{-1}(e_l)\Bigr)vS^2(e_i)S^3(e_k)f_le_j\nonumber\\
&=& \nu^5   \,\sum_{jl}\,\sum_{ik}\,\mu_D\Bigl(vA\,S^{-1}(f_j)\bigl(f_i
f_k\bigr) S^{-1}(e_l)\Bigr)vS^2\bigl(e_iS(e_k)\bigr)f_le_j\nonumber\\
^{by(\ref{dual3})}&=& \nu^5
\sum_{jl}\,\mu_D\Bigl(vA\,S^{-1}(f_j)S^{-1}(e_l)\Bigr)
vf_le_j
\eea
We readily identify the last equation with the right hand side of
(\ref{ststst}). This completes the proof.
\hfill $\Box$

\medskip
The $\cal S$ matrix was originally defind as an element in the $End$ -set of
the coend of the representation category. As a map on $\dA$  it
 therefore intertwines the $ad^-$ - action ( see  (\ref{gam2})) of the algebra
on itself. (This property can also be inferred directly from Lemma \ref{veup}.
{\em 1.)}.) The same is true for multiplications with central elements as for
example for $\cal T\,$. Hence the center $Z(\dA)\,$ - which is the invariance
of
the $ad^-$ - action - is an invariant subspace of both operators.
It follows immediately from (\ref{BrSexp}) that the restriction of $\Gamma$ to
the center is the usual antipode $S$ and thus involutive.

We summarize these observations and the relations found in (\ref{gam2}),
(\ref{morelex}), and (\ref{ststst}) in the following theorem:

\btm
Suppose $\dA$ is the double of a finite dimensional Hopf algebra. Assume that
$\dA$ admits a balancing and let $\nu$ and $\omega$ be as in Proposition
\ref{moduli} . Furthermore, let $\cal T$ be the multiplication with $v$, and
$\cal S$ and $\Gamma$ be defined as in (\ref{ffcS}) and (\ref{BrSexp}),
respectively. Then
\ben
\item The generators define a projective representation of the
mapping class group $\;{\cal D}\;:=$ $\pi_o\bigl( \diff (T,D)\bigr)$
 of torus maps fixing a disk
with the following relations:
\bea\label{uuu}
{\cal S}^2\,&=&\,\omega\,\Gamma^{-1}\qquad\qquad\quad
{\cal T}\Gamma\,=\,\Gamma{\cal T}\\
\Bigl({\cal S}{\cal T}\Bigr)^3\,&=&\,\nu^3\,\Gamma^{-2}\,=\,\nu^3ad^-(v)
\eea
\item The maps $\cal S$ and $\cal T$ stabilize the center $Z(\dA)\,$.
The restrictions $\bor{\cal S}$ and $\bor{\cal T}$ satisfy
\beq\label{ffff}
\bigl(\bor{\cal S}\,\bor{\cal T}\bigr)^3\,=\,\nu^3\;\qquad\bor{\cal
S}^2\,=\,\omega\,S^{\pm 1}\;\qquad {\cal T}S\,=\,S{\cal T}
\eeq
where $\cal S$ is the involutive map given by the restriction of the antipode
to the center.
\een
\etm

\medskip
The relations in (\ref{ffff}) show that  $\bor{\cal S}$ and $\bor{\cal T}$
define a projective representation of $SL\bigl(2,{\bf Z}\bigr)\,$. The
normalization of the $\cal S$-operation
was defined by the canonical normalization of $\mu_D\,$. For the computation of
topological invariants it is often more convenient to have a normalization for
which the operators are inverted if we invert the
braided structure. For a given balancing $k$ let  $\bor{\cal S}'\,$ and
$\bor{\cal T}'\,$ be the analogous  operators defined with respect to
$R'\,=\,\tau\bigl(R^{-1}\bigr)\,$. Then as $u'=\hat u$ we have that
$\bor{\cal T}'=\bor{\cal T}^{-1}\,$ is already correctly normalized. A
computation similar to the one in
Proposition \ref{Smatrix} yields
$$
\bor{\cal S}'\,=\,\omega\,\bor{\cal S}^{-1}\;.
$$
Thus it is the matrix ${\cal S}_{\bdot} := \nu^2 \bor{\cal S}^{-1}\,$ which
inverts under inversion of
the braided structure. For these generators we have the relations:
\beq\label{bestrel}
{\cal S}_{\bdot}^4=1 \qquad\qquad\qquad \bigl({\cal S}_{\bdot}\bor{\cal
T}\bigr)^3=\nu^{-3}{\cal S}_{\bdot}^2
\eeq
Comparing (\ref{bestrel}) to relations in [T] and [RT] we find that the
projective phase $c\,$ of the
functor $\Phi$ in (\ref{projfun}) for a universal TQFT over a double $\dA\,$ is
given by:
$$
c\;=\;{\nu}^{-3}\qquad.
$$

\bigskip

\bigskip

\setcounter{chapter}{3}
\section*{3.The Relation of Universal and Semisimple TQFT's: An Example }

In this section we shall analyze the proposed representation of the mapping
class group $\cal D$ of the punctured torus explicitly in the example of the
double
of the quantum-$sl_2$-Borel algebra $B_q\,$.

\noindent
\underline{1.) The Algebra $D(B_q)$\,} : Let $q$ be a primitive $l\,$-th root
of unity
where $l=2m+1$, $m \in{\bf Z}_{\geq 1}\,$. We denote by $B_q$ the
 Hopf algebra with generators $e,\,k^{\pm 1}$ and relations:
\begin{equation}\label{Bqdef}
\begin{array}{lll}
kek^{-1}\,=\,qe\,, \quad &  k^l\,=\,1  \quad & \;e^l\,=\,0 \\
\Delta (k)\,=\,k\otimes k\quad & \Delta(e)\,=\,e\otimes 1 \,+\,k^2\otimes e &\\
S(e)\,=\,-k^{-2}e\quad & \, S(k)\,=\,k^{-1} \quad & \epsilon(e)=0 \quad  \,
\epsilon(k)=1 \;.\\
\end{array}
\end{equation}

\noindent
As  PBW-basis for $B_q$  we choose $ e^n k^j\,$ with $n=0,\ldots ,l-1\,$ and $j
\in  {\bf Z}/l$.

\noindent
The left {\it cointegral} of $B_q$ is given by
\begin{equation}\label{coint}
 x\,=\, \bigl(\sum_{j=0}^{l-1} k^j\bigr)e^{l-1}
\end{equation}
and the left {\it integral} with normalization $m(x)=1$ is
\begin{equation}\label{int}
m(e^nk^j)\,=\,q^2\delta_{j,2}\delta_{n,(l-1)}\;.
\end{equation}
The {\it moduli} of these integrals are easily found to be
\beq\label{mod}
a\,=\,k^2\qquad\quad
{\rm and}\qquad\quad
\alpha(k)\,=\,q,\quad \alpha(e)=0\;
\eeq
\begin{equation}\label{omega}
{\rm so\; that}\qquad\qquad\qquad\qquad\omega = q^2 \;.\qquad\qquad\qquad\qquad
\end{equation}
Since we assumed $l$ to be odd we can
choose as generators of the dual algebra $B_q^*$ the modulus $\alpha$ and
the linear form $f$ defined by $<f,e^nk^j>=\delta_{n,1}$.
The following relations together with those in (\ref{Bqdef}) can be used as  a
definition for the double $D(B_q)$ containing $B_q$ and $B_q^*$ with opposite
comultiplication:
\begin{equation}\label{DBdef}
\begin{array}{lll}
\alpha f \alpha^{-1}\,=\,q^2f & \alpha^l=1 & f^l=0 \\
\alpha e \alpha^{-1}\,=\,q^{-2}e & kfk^{-1}\,=\,q^{-1}f&\\
ef -fe\,=\,\alpha - k^2&&\\
\Delta(\alpha)=\alpha\otimes\alpha&\Delta(f)=f\otimes\alpha + 1\otimes f &\\
\end{array}
\end{equation}
We shall sometimes refer to the {\it $\bf Z$ - gradation}  of $D(B_q)$ which is
defined on the generators by
$ gr(e) =+1\,$, $gr(k)=0\,$,
$gr(f)=-1\,$, and  $gr(\alpha)=0\,$.
The universal $\cal R$ -matrix of this algebra is
\begin{equation}\label{Rmat}
{\cal R}\,=\, \bigl(\sum_{n=0}^{l-1}\frac {q^{-\frac {n(n-1)} 2}}
{[n]!}e^n\otimes f^n\bigr)\bigl(\frac 1 l \sum_{i,j \in  {\bf Z}/l}
q^{-ij}k^j\otimes \alpha^i \bigr)
\end{equation}
Here $[n]!\,=\,[n][n-1]...[1]$ with $[n]\,=\,\frac {q^n-q^{-n}}{q-q^{-1}}$.
If we compute the expressions in (\ref{defbet}) and (\ref{defborbet}) for the
integrals in (\ref{coint}) and (\ref{int}) we obtain the following isomorphisms
between $B_q$ and $B_q^*$ :
\begin{equation}\label{betal}
\beta_l(k^je^n)\,=\,\frac
{q^{-\frac{(l-1-n)(l-2-n)}2}}{[l-1-n]!}\,f^{l-1-n}\,\frac 1 l \sum_{i\in {\bf
Z}/l}q^{i(j-2)}\alpha^{i+1}\;.
\end{equation}
and
\begin{equation}\label{betar}
\bor{\beta_l}(\alpha^if^n)\,=\,(-1)^n[n]! q^{-\frac{n(n+3)} 2 -
2i}e^{l-1-n}\sum_{j\in {\bf Z}/l}q^{j(i-1)}k^j\;.\\
\end{equation}
As an associative algebra $D(B_q)$ is isomorphic to the product ${\bf C}[{\bf
Z}/l]\otimes
U_q(sl_2)\,$, where the central group algebra ${\bf C}[{\bf Z}/l]$ is generated
by
\begin{equation}\label{defzd}
z\,:=\,\alpha^{-m}k
\end{equation}
The generators of the $U_q(sl_2)$ factor are defined by
\begin{equation}\label{genUsl2}
E:=z^{-1}e, \quad F:=-f, \quad K:=\alpha^{m}k
\end{equation}
and obey the relations
\begin{equation}\label{relUsl2}
KEK^{-1}\,=\,q^2E, \quad KFK^{-1}\,=\,q^{-2}F,
\end{equation}
\begin{equation}
EF-FE\,=\,K-K^{-1}\;.\\
\end{equation}

\bigskip

\noindent
\underline{2.) The  Center of $D(B_q)$ } :
Thanks to the above decomposition the center of $D(B_q)$ is given by  ${\bf
C}[{\bf Z}/l]\otimes {\cal V}$ where ${\cal V}$ is the center of
$U_q(sl_2)\,$.
In order to give a description of ${\cal V}$ it is convenient to introduce the
projections
\begin{equation}\label{piK}
\pi_j(K)\,=\,\frac1 l \sum_{i\in {\bf Z}/l}q^{2ij}K^i\qquad j\in {\bf Z}/l
\end{equation}
on the eigenspaces of $K$ with eigenvalue $q^{-2j}$. Furthermore we introduce
the projections
\begin{equation}
T_j\,=\,\sum^{l-1-j}_{s=j+1}\pi_s(K) \qquad j=0,\ldots,m-1\;.
\end{equation}
The standard quadratic Casimir of $U_q(sl_2)$ is given by:
\begin{equation}\label{defX}
X\,=\,EF+ \frac{qK^{-1}+q^{-1}K}{q-q^{-1}}\;.
\end{equation}
The trivially graded part $U^o$ of $U_q(sl_2)$ (gr(E)=1,gr(F)=-1,gr(K)=0) is a
free module over the ring ${\bf C}[K]$ with basis $\{X^j\}_{j=0,...,l-1}$ and
the
minimal equation for $X$ is:
\begin{equation}\label{chaX}
\prod_{j=0}^{l-1}\bigl(X-b(j)\bigr)\,=\,0\;,
\end{equation}
where the roots
\begin{equation}\label{root}
b(j)=b(l-1-j):=\,\frac {q^{(2j+1)}+q^{-(2j+1)}}{q-q^{-1}}
\end{equation}
are of order two for $j=0,\ldots,(m-1)$ and of order one for $j=m$.

Using the polynomials
\begin{equation}\label{phi}
\phi_j(X)\,=\,\prod_{0\leq s\leq (l-1) : b(s)\neq b(j)}\bigr(X-b(s)\bigl)\qquad
\; j=0,\ldots,m
\end{equation}
of order $(l-2)$ for $j<m$ and of order $(l-1)$ for $j=m$ we can define the
idempotents and nilpotents associated to $X$:
\begin{equation}\label{defZ}
\begin{array}{rll}
P_j\,&=\,\frac 1
{\phi_j(b(j))}\phi_j(X)\,-\,\frac{\phi_j'(b(j))}{\phi_j(b(j))^2}
\bigl(X-b(j)\bigr) \phi_j(X)\qquad &j=0,\ldots,m\\
&&\\
N_j\,&=\,\frac 1 {\phi_j(b(j))}\,\bigl(X-b(j)\bigr) \phi_j(X)\qquad
&j=0,\ldots\,(m-1)\\
&&\\
N_j^+\,&=\,T_j\,N_j\qquad N_j^-\,=(1-T_j)\,N_j\\
\end{array}
\end{equation}

For example a general polynomial $\Psi(X)$ in $X$ is expressed in terms of
$P_j$ and $N_j$ by the formula:
\begin{equation}\label{expZ}
\Psi(X)\,=\,\sum_{j=0}^m \Psi(b(j))P_j\,+\,\sum_{j=0}^{m-1}\Psi'(b(j))N_j\quad
{}.
\end{equation}

The normalizations in (\ref{defZ}) can be evaluated explicitly using
\begin{equation}\label{phib}
\begin{array}{rl}
 \phi_j(b(j))\,&=\,\frac {l^2}{(q-q^{-1})^l} \frac 1 {[d_j^{\pm}]^2}\\
 \phi'_j(b(j))\,&=\,-\frac{l^2}{(q-q^{-1})^{(l+1)}} \frac
{[2d_j^{\pm}]}{[d_j^{\pm}]^5}\qquad\qquad {\rm for } \quad j=0,\ldots,(m-1)\\
\phi_m(b(m))\,&=\,\frac {l^2}{(q-q^{-1})^{(l-1)}}
\end{array}
\end{equation}

The center of the quantum algebra does not only contain the subalgebra
generated by $X$ but also the above combinations of nilpotents with the
weight-projectors $T_j$. More precisely, we have the following lemma:
\begin{lemma}\label{center}
The center, denoted by $\cal V$ of $U_q(sl_2)$ is the $(3m+1)$-dimensional
algebra with basis
$\{P_i,\,N^{\pm}_j\,:\,i=0,\ldots, m\,;j=0,\dots,m-1\,\}$ and products:
\begin{equation}\label{PNrel}
\begin{array}{rll}
P_iP_j\,&=\,\delta_{ij}P_j&\\
P_iN^{\pm}_j\,&=\delta_{ij}N^{\pm}_j&\\
N^{\pm}_iN^{\pm}_j\,&=N^{\pm}_iN^{\mp}_j&=0\qquad .\\
\end{array}
\end{equation}
\end{lemma}

{\it Proof :} We use the fact that every element $y$ in the trivially graded
 part $U_o$ has a unique presentation:
$$
y\,=\,\sum_{s\in {\bf Z}/l}\pi_s(K)p_s(X)
$$
wher the $p_s$ are polynomials of order smaller than $l$. The condition that
$y$ commutes with $E$ is then:
$$
\sum_{s\in {\bf Z}/l} \pi_s(K)\bigl(p_s(X)-p_{s-1}(X))\;\in\, {\cal I}
$$
Here we denote the ideal ${\cal I} \,=\,\{ y\in U^o : Ey=0\}\,$. It is clear
that $\cal I$ is generated by
$$
E^{l-1}F^{l-1}\,=\,\sum_{s\in {\bf Z}/l} \pi_{s+1}(K)\prod_{j\in {\bf Z}/l,\;
j\neq s}\Bigl(X-b(j)\Bigr)\;.
$$
The polynomials in $X$ occurring in this sum are proportional to the nilpotents
and
idempotents defined in (\ref {defZ}). The ideal $\cal I$ is therefore spanned
by the elements
$$
 \pi_{s+1}(K)N_s,\,\,s=0,...,m-1\;{\rm  and}\qquad \pi_{m+1}P_m\,.
$$
Solving the recursion for the $p_j$'s we find that the center is generated by
the elements in (\ref{defZ}). The commutation with $F$ yields exactly the same
conditions. Linear independence
of these generators can be shown by choosing special representations of $X$.

\hfill$\Box$

\noindent
\underline{3.) Canonical Elements and Balancing : }

 The canonical, group like  element, $g\,$, from (\ref{defg}) implementing the
fourth order of the antipode is obtained from the equations for the moduli:
\begin{equation}\label{radf}
g\,=\,\alpha k^{-2}\,=\,K^{-2}\;.
\end{equation}
For odd $l$ this element has precisely one square root in the group like
elements,
\begin{equation} \label{sqrg}
\sqrt g \,=\, K^{-1}\;
\end{equation}
so that we have uniqueness of balancing. The fourth root of one associated to
this by (\ref{nu}) is
\beq
\nu\,=\,q^{-m}
\eeq
The canonical element $u\,=\,S(R^{(2)})R^{(1)}\,$ can be expressed as
the following product of commuting elements
\begin{equation}\label{equ}
u\,=\,u_Z u_K u_o
\end{equation}

$$
{\rm where}\qquad\qquad\qquad u_Z\,=\,\frac {\gamma_q} {\sqrt l} \sum_{i\in
{\bf Z}/l} q^{-mi^2}z^i
\qquad\qquad\qquad u_K\,=\,\frac {{\gamma_q}^{-1}} {\sqrt l}\sum_{i\in {\bf
Z}/l} q^{mi^2}K^i
$$
$$
{\rm and }\qquad\qquad\qquad u_o\,=\,\sum_{n=0}^{l-1}
\frac{q^{\frac{n(n+3)}2}}{[n]!} K^nF^nE^n
$$

Here we denote the Gauss sum
$
\gamma_q\,:=\,\frac 1 {\sqrt l} \sum_{j=0}^{l-1}q^{mj^2}\;,
$
which is a phase for odd $l$ and can be evaluated explicitly ( see e.g.
[L]).
 The unique ribbon element $v$ can be written as a product of
an element in the ${\bf C}[\bf Z/l]$ factor and an element in the
$U_q(sl_2)$-factor of the algebra:
$$
v\,=\,u_Zv_o
$$
where
\begin{equation}\label{bal}
v_o\,=\,Ku_Ku_o
\end{equation}
If we  denote by ${\cal T}_Z$ , ${\cal T}_o$ and ${\cal T}$ the linear
operators on
${\bf C}[\bf Z/l]$, $U_q(sl_2)$ and $D(B_q)$ defined by multiplication with
$u_Z$, $v_o$ and $v$ respectively, this implies
\begin{equation}\label{Tfac}
{\cal T}\,=\,{\cal T}_Z \otimes {\cal T}_o
\end{equation}

We have the following expression for the central element $v_o\,$ in terms of
the basis given in Lemma \ref{center} :
\begin{lemma}\label{ribbond}
The central ribbon element $v_o \,\in\,U_q(sl_2)$  has the Jordan decomposition
\begin{equation}
v_o\,=\,q^mP_m
\,+\,\sum_{j=0}^{m-1}q^{2j(j+1)}\Bigl(P_j+\frac{d^+_j}{[d^+_j]}N^+_j+
\frac{d^-_j}{[d^-_j]}N^-_j\Bigr)\;,
\end{equation}
Here the basis elements of $\cal V$ are the same as in (\ref{defZ}) and the
numbers \linebreak \mbox{$d^{ \pm}_j=1,...,l-1$}, are defined for $j=0,...,m-1$
by
$$
d^+_j\,:=\,2j+1\qquad
d^-_j\,:=\,l-(2j+1)\;.
$$
\end{lemma}
{\it Proof :} The computation of these coefficients is most conveniently done
by
multiplying the expression for $v_o$ obtained from (\ref{equ}) by a weight
projector $\pi_s(K)$. The result can be expressed in terms of a polynomial
$\Psi_s$ of the quadratic Casimir $X$:
\begin{equation}\label{Psi}
\pi_s(K)v_o\,=\,\pi_s(K)\Psi_s(X)\;,
\end{equation}
where
$$
\Psi_s(X)\,=\,\sum_{n=0}^{l-1} \frac {q^{\frac
{n(n+3)}2}}{[n]!}q^{2a(a-n-1)}\prod_{i=l-n}^{l-1}\bigl(X-b(i+s))\;.
$$
{}From the general expansion (\ref{expZ}) we see that the coefficient of $P_j$
is given by $\Psi_s(b(j))$ for any $s$ and the coefficients of $N^+_j$
and $N^-_j$ are given by $\Psi_s'(b(j))$  where \linebreak
\mbox{$s=j+1,...,l-1-j$} and
\mbox{$s=-j,...,j$} respectively. For a choce of $s$ with $b(s-1)=b(j)$ we can
avoid one summation in the expressions for $\Psi_s$ and $\Psi_s'\,$.
In order to evaluate the remaining sum for $\Psi'$ we invoke the
partition identity for $t$ with $t^i\neq 1$ for $i=1,...,d\;$:
\begin{equation}\label{part1}
\frac d {1-t^d}\,=\,\sum_{n\geq 1}\frac 1
{1-t^n}\prod_{i=1}^{n-1}\bigl(1-t^{(d-i)}\bigr) \;.
\end{equation}
\hfill$\Box$

{\it Remark :} From the observation that the coefficients should be independent
of the
choice of the weight $s$ we are led to new partition identities.
For example in the computation of $\Psi_s$ we find the formula:
$$
t^{AB}\,=\,\sum_{n=0}^{min(A,B)}\,\prod_{i=0}^{n-1}\frac{(t^A-t^i)(t^B-t^i)}{t^i(t^{(i+1)}-1)}\;.
$$

\bigskip
\noindent
\underline{4.) The $SL(2,{\bf Z})$-Action on the Center of $D(B_q)$ : }
We use the formula obtained in (\ref{Smatrix}) to give the explicit action of
${\cal S}\,$
on $D(B_q)\,$. Together with $\cal T$ defined by multiplication with the ribbon
element
this yields a representation of the mapping class group $\cal D$ on $D(B_q)\,$.
If we insert the expressions for the integrals  from (\ref{betal}) and
(\ref{betar}) the action of $\cal S$ can be immediately written if
we use both PBW bases  $ k^je^nf^p{\alpha}^s $  and $ {\alpha}^sf^ne^Pk^j $
as:
$$
{\cal S}(k^je^nf^p{\alpha}^s)\,=\,\frac { (-1)^n[p]!}{[l-1-n]!}\,q^{\bigl(\frac
{(n+1)(n+2)}2 +(n+1)j
+\frac{p(p-1)}2\bigr)}\times\hphantom{xxxxxxxxxxxxxxxxxxxxxx}
 $$
\begin{equation}\label{S1}
\qquad\qquad \Bigl(\frac 1 l \sum_{i\in \bf Z/l}q^{-ij}\alpha^i\Bigr)
f^{(l-1-n)}e^{(l-1-p)}\Bigr(\sum_{i\in \bf Z/l}q^{i(s+p)}k^{-i}\Bigl)\;\quad.
\end{equation}
A similar formula was obtained in [LyM].
It is immediate from the above form that the $\cal S$ - matrix  preserves the
gradation $n-p\,\in \bf Z$ of a basis element. Given that the balancing element
is trivially
graded and acts by multiplication it follows that the $\cal D$ - representation
on $ D(B_q) $ decomposes into a direct sum of the $2l-1$ spaces corresponding
each gradation.

Clearly, the category from which $\cal S$ is obtained is the tensor product of
the representation category of $U_q(sl_2)$ and ${\bf C}[\bf Z/l]$ as an abelian
category. Also, since the balancing element and hence the monodromy can be
factorized into a product of invertible elements from either algebra
the $\cal S$ - matrix has to factorize too. More precisely we define the
following isomorphisms on $C[\bf Z/l(z)]$,
\begin{equation}\label{SZ}
{\cal S}_Z(z^n)\,:=\,\frac 1 {\sqrt l}\sum_{j\in\bf Z/l} q^{-jn}z^j \quad .
\end{equation}
and on $U_q(sl_2)$
$$
{\cal S}_o(K^jE^nF^p)\,:=\,\frac { (-1)^p[p]!}{[l-1-n]!}\,q^{\bigl(\frac
{(n-p)(n-p+1)}2 +j(2n+1-p)+1\bigr)}\times\hphantom{xxxxxxxxxxxxxxxxxxxxxx}
$$
\begin{equation}\label{SU}
\qquad\qquad\Bigl(\frac 1 {\sqrt l} \sum_{k\in\bf
Z/l}q^{k(j-n)}K^k\Bigr)F^{(l-1-n)}E^{(l-1-p)}\qquad.
\end{equation}
Using the isomorphism  $D(B_q)\,\cong\,C[{\bf Z}/l]\otimes
U_q(sl_2)$ defined by the change of basis in (\ref{defzd}) and (\ref{genUsl2})
we can now write the $\cal S$ -matrix in the form:
\begin{equation}\label{Sfac}
{\cal S}\,=\,{\cal S}_Z\otimes {\cal S}_o\qquad\quad.
\end{equation}

Together with (\ref{Tfac}) this shows that the representation of $\cal D$
on $D(B_q)$ is given by the tensorproduct of two projective representations of
 $\cal D$. Since $C[{\bf Z}/l]$ is central in  $D(B_q)$ we expect the
representation generated by ${\cal T}_Z$ and ${\cal S}_Z$ to factor through a
projective representation of $SL(2,\bf Z)$. In fact we easily verify the
following relations
\begin{equation}\label{SLrelZ}
{{\cal S}_Z}^2 {\cal T}_Z=\,{\cal T}_Z{{\cal S}_Z}^2\;,\qquad\quad {{\cal
S}_Z}^2(z^n)=z^{-n}\;,\qquad\quad
\bigl({\cal S}_Z{\cal T}_Z\bigr)^3=\,\gamma_q\id
\end{equation}


 It is clear that the action of ${\cal S}_o$ on $U_q(sl_2)$ preserves the
gradation in the same way as the action of ${\cal S}$ on $D(B_q)$. For
example the restriction on the highest $l-1$ graded subspace defines for each
$g_o\,\in\,\cal D$ by
$$
g_o(aKE^{l-1})\,=\,\hat g (a)KE^{l-1}
$$
an action $\hat g$ on an element $a$ in the group algebra ${\bf C}[{\bf Z}/l]$
generated by $K$ . It factors into an  $SL(2,{\bf Z})$
 representation and is equivalent to the one defined previously by ${\cal T}_Z$
and ${\cal S}_Z$ with $q$ replaced by $q^{-1}$.

\medskip
In the following we shall focus on the $0$ -graded part $U_o$ of $U_q(sl_2)$
from where we wish to compute the restrictions to the center.
 We determine explicitely the $(3m+1)$ - dimensional representation
matrices of $SL(2,\bf Z)$ which we obtain by restricting the action of
$\cal D$ onto the center $\cal V$ of $U_q(sl_2)\,$. We choose the basis
as in Lemma \ref{center}  in the order
$P_0,\,N^+_0,\,N^-_0,\,P_1,\ldots,\,N^-_{(m-1)},\, P_m\;$. On the subspace
spanned by $P_j,\,N^+_j,\,N^-_J$ we define the Jordan bloc:
$$
\tau_j\,:=\;q^{2j(j+1)}\left[\begin{array}{ccc}1 &0&\;0\\ \frac
{d^+_j}{[d^+_j]}&1&\;0\\ \frac
{d^-_j}{[d^-_j]}&0&\;1\end{array}\right]\qquad{\rm for}\;j=0,\ldots, (m-1)\;.
$$
Then it is obvious from the formula in Lemma \ref{ribbond} that the ${\cal
T}_o$ matrix defined by multiplication of $v_o\,$ is given by the direct sum:
\begin{equation}\label{Tmat}
{\cal T}_o\,=\,\tau_0\oplus\tau_1\oplus\ldots\oplus\tau_{(m-1)}\oplus q^m
\end{equation}

The restriction of ${\cal S}_o$ to the center is much more complicated and
will be dealt with in the rest of this section. It involves finding
transformations from the idempotents and
nilpotents given in Lemma \ref{center} to polynomials in $X$ and $K$, to the
standard PBW basis of $U_q(sl_2)$  and backwards. The transformations between
the center
and expressions in $X$ and $K$ can be obtained from the relations given in
(\ref{defZ}) and (\ref{expZ}). In order to reexpress polynomials in
$X$ and $K$ in terms of the basis $K^jE^nF^n$ and conversely we need the
following two technical lemmas. A special case of the first lemma we used
already in the computation of the center. The proof is straightforward.
\begin{lemma}\label{PBWtoKX}
Let $X$ be the quadratic Casimir defined in (\ref{defX}) and set
\begin{equation}\label{defQ}
Q_j\,=\,\frac {q^{(2j+1)}K^{-1}+q^{-(2j+1)}K}{q-q^{-1}}
\end{equation}

then the following relations hold
\begin{eqnarray}
E^jF^j\;=\;\prod_{s=0}^{j-1}\bigl(X-Q_s\bigr)\label{EFtoX}\\
F^jE^j\;=\;\prod_{s=l-j}^{l-1}\bigl(X-Q_s\bigr)\label{FEtoX}\;.
\end{eqnarray}
\end{lemma}

Before we give the converse transformations let us state the following
identity for general polynomials
\begin{lemma}\label{polyn}
Suppose $\lambda_0,\ldots,\lambda_N$ is an ordered set of roots and
$0 \leq a_1<a_2 \ldots <a_k \leq N$ are an ordered set of $k$ indices then
we have
\begin{equation}\label{polyid}
\prod_{j=0,\;j\noin \,\{a_j\}}^N\!\!\!\bigl(X-\lambda_j\bigr)\,=\,\sum_{0\leq
s_1 <s_2
<\ldots<s_k\leq
N}\,\prod_{i=0}^{s_1-1}\!\bigl(X-\lambda_i\bigr)\prod_{i=s_1+1}^{s_2-1}\!\!\!\bigl(\lambda_{a_1}-\lambda_i\bigr)\ldots\prod_{i=s_k+1}^N\!\!\!\bigl(\lambda_{a_k}-\lambda_i\bigr)
\end{equation}

reexpressing a polynomial with omitted roots in terms of polynomials with
consecutive roots.
\end{lemma}

Here an empty product is meant to be $1$. The proof is a straight forward
induction which is most conveniently done by assuming the statement for
for all $(k,N')$ with $k'< k\,$ or $\,k'=k$ and $N'\leq N$ and proving it for
$\,k'=k$ and $N'=N+1$ thus for all pairs with $k'=k$. This is followed by an
induction in $k$. If we combine Lemma \ref{PBWtoKX} and Lemma \ref{polyn}
we arrive at the following formula for the polynomials defined in (\ref{phi})
\begin{lemma}\label{phiPBW}
Let $\phi_k(X)$ be the polynomials in the quadratic Casimir $X$ as defined in
(\ref {phi}), $\pi_s(K)$ the projector from (\ref{piK}) and $b(j)$ as in
(\ref{root}). Then

\begin{equation}\label{phitoEF}
\pi_t(K)\bigl(X-b(k)\bigr)\phi_k(X)\,=\,\sum_{j=0}^{l-1}\,\prod_{i=j+1}^{l-1}\!\bigl(b(k)-b(i+t)\bigr)\;\pi_t(K)E^jF^j\;,\hphantom{xxxxxxx}
\end{equation}
\begin{equation}\label{XphitoEF}
\pi_t(K)\phi_m(X)\,=\,\sum_{j=0}^{l-1}\,\prod_{i=j+1}^{l-1}\!\bigl(b(m)-b(i+t)\bigr)\;\pi_t(K)E^jF^j\;,\hphantom{xxx}
\end{equation}
\begin{equation}\label{phimtoEF}
\pi_t(K)\phi_k(X)\,=\,\sum_{j=0}^{l-2}\sum_{s=j+1}^{l-1}\;\prod_{\stackrel{i=j+1}{i\neq s}}^{l-1}\!\bigl(b(j)-b(i+t)\bigr)\;\pi_t(K)E^jF^j\;.
\end{equation}
\end{lemma}

{\it Proof } : We apply Lemma \ref{polyn} to the situation where $N=l-1$, $X$
is the
quadratic Casimir and the roots $\lambda_j$ are replaced by the elements $Q_j$
defined in  (\ref{defQ}). The polynomials with consecutive roots on the right
hand side of (\ref{polyid}) are precisely those in (\ref{EFtoX}). Thus for
$k=1,2$ we obtain the specializations:
\begin{equation}\label{Q1}
\prod_{\stackrel{j=0}{j\neq a}}^{l-1}\!\!\bigl(X-Q_j\bigr)\;
=\;\sum_{j=0}^{l-1}\prod_{s=j+1}^{l-1}\!\!\bigl(Q_a-Q_s\bigr)\,E^jF^j\;.
\hphantom{xxxxxxxxxxxxxxx}
\end{equation}
and
\begin{equation}\label{Q2}
 \prod_{\stackrel {j=0}{j\neq a,b}}^{l-1}
\!\bigr(X-Q_j\bigl)\;=\;\sum_{j=0}^{l-1}\sum_{s=j+1}^l\,\prod_{i=j+1}^{s-1}\!\!\bigl(Q_p-Q_i\bigr)\prod_{i=s+1}^l\!\!\bigl(Q_b-Q_i\bigr)\;E^jF^j\;.
\end{equation}
Notice that the polynomials $\phi_k(X)$ and $(X-b(k))\phi_k(X)$ are obtained
from (\ref{chaX}) by omitting one or two roots. If we multiply equation
(\ref{Q1}) and(\ref{Q2}) with the projector $\pi_t(K)$ for suitable choices of
$a$ and $b$ we obtain these polynomials on the left hand side. The identities
(\ref{phitoEF})  - (\ref{phimtoEF}) follow using
$\pi_t(K)Q_j\,=\,\pi_t(K)b(j+t)\,$.

\hfill $\Box$

Lemma \ref{phiPBW} puts us now in the position to determine the action of
${\cal S}_o$ on the polynomials $\phi_k(X)$ and
$\bigl(X-b(k)\bigr)\phi_k(X)\,$.

Insertion into (\ref{SU}) yields for \mbox{$k=0,\ldots,(m-1)\,$}:
\begin{equation}\label{Sonpol1}
{\cal S}_o\Bigl(\pi_t(K)\bigl(X\!-\!b(k)\bigr)\phi_k(X)\Bigr)\,=\,\sum_{b\in\bf
Z/l}\pi_b(K)\Delta^{tk}_b(X)\;\;,\hphantom{xxxxxxx}
\end{equation}
\begin{equation}\label{Sonpol2}
{\cal S}_o\Bigl(\pi_t(K)\phi_k(X)\Bigr)\,=\,\sum_{b\in\bf
Z/l}\pi_b(K)\Gamma^{tk}_b(X)\;\;,
\end{equation}
\begin{equation}\label{Sonpol3}
{\cal S}_o\Bigl(\pi_t(K)\phi_m(X)\Bigr)\,=\,\sum_{b\in\bf
Z/l}\pi_b(K)\Delta^{tm}_b(X)\;\;.
\end{equation}

\noindent
The polynomials $\Delta^{tk}_b(X)$ and $\Gamma^{tk}_b(X)$ are defined by
\begin{equation}\label{delta}
\Delta^{tk}_b(X)\,:=\,\hphantom{xxxxxxxxxxxxxxxxxxxxxxxxxxxxxxxxxxxxxxxxxxxxxxxxxxxxxx}
\end{equation}
$$
\frac q {\sqrt l} \sum_{j=0}^{l-1}\,\frac
{(-1)^j[j]!}{[l-1-j]!}\,q^{(j+2b)(j+2t+1)}\prod_{i=j+1}^{l-1}\!\bigl(b(k)-b(i+t)\bigr)\prod_{i=j+1}^{l-1}\!\bigl(X-b(i+b)\bigr)\;,
$$
and
\begin{equation}\label{gamma}
\Gamma^{tk}_b(X)\,:=\,\hphantom{xxxxxxxxxxxxxlxxxxxxxxxxxxxxxxxxxxxxxxxxxxxxxxxxxxxxxxx}
\end{equation}
$$
\frac q {\sqrt l}
\sum_{j=0}^{l-2}\sum_{s=j+1}^{l-2}\,\frac{(-1)^j[j]!}{[l-1-j]!}\,q^{(j+2b)(j+2t+1)}
\prod_{\stackrel{i=j+1}{i\neq s}}^{l-1}
\!\bigl(b(k)-b(i+t)\bigr)\prod_{i=j+1}^{l-1}\!\bigl(X-b(i+b)\bigr)
$$
The action of ${\cal S}_o$ on $\cal V\,$ is now easily obtained by
summing equations (\ref{Sonpol1}) to (\ref{Sonpol3}) over an appropriate range
of $t$ and
combining them in (\ref{defZ}). The result contains products of projections
$\pi_b(X)$ with polynomials of the form $\sum_t \Delta^{kt}_b(X)$ and $\sum_t
\Gamma^{kt}_b(X)\,$ .
The polynomials can be expanded for every weight $b$ into the basis $\cal V$
defined in Lemma\ref{center}; the coefficients of $P_p$ and $N_p$ are obtained
by the values and values of the derivatives of these polynomials at $X=b(p)\;$.

By general construction the ${\cal S}_o$ - matrix has to map $\cal V\,$ to
itself. As in the remark following Lemma \ref{ribbond} this can be used to
produce new families of
partition identities.

In order to find the matrix coefficients of the  $SL(2,\bf Z)$ representation
we need the following quantities:

\beq
\eta\bigl( d_A , d_B \bigr)\, := \hphantom{xxxxxxxxxxxxxxxxxxxxx}
\hphantom{xxxxxxxxxxxxxxxxxxxxxxxxxxxxxxxxx}\label{defeta}
\eeq
$$\,[d_A]^2
\frac{(q-q^{-1})^l}{l^2}\,\,\sum^{min(d_B,\,d_A-1)}_{j=1}\sum^{d_A}_{s=j+1}\,\,[l-1-j]!\frac{(-1)^j}{[j]}\,q^{(j-d_B)(j+1+d_A-2s)}\,\times\qquad
$$
$$
\hphantom{xxxxxxxxxxxxxxxxxxxxxxxx}
\prod_{i=1}^j\!(q-q^{-1})[s-i][d_A-s+i]\,\prod_{i=1}^{j-1}\!\Bigl(q^{(d_B-i)}-
q^{-(d_B-i)}\Bigr)\;,
$$
\begin{equation}\label{defmu}
\mu(d_A,d_B)\,:=\hphantom{xxxxxxxxxxxxxxxxxxxxxxxxxxxxxxxxxxxxxxxxxxxxxxxxxxxxxx}
\end{equation}
$$\,[d_A]^2 \frac{(q-q^{-1})^l}{l^2}\,\,\sum_{s\in\bf
Z/l}\sum^{d_B}_{j=1}\sum^{j}_{r=1}\,\,[l-1-j]!\frac{(-1)^j}{[j]}\,q^{(j-d_B)(j+1+d_A-2s)}\,\times\qquad
$$
$$
\hphantom{xxxxxxxxxxxxxxxxxxxxxxxx} \prod_{\stackrel{i=1}{i\neq
r}}^j\!(q-q^{-1})[s-i][d_A-s+i]\, \prod_{i=1}^{j-1}\!\Bigl(q^{(d_B-i)}-
q^{-(d_B-i)}\Bigr)
$$

and
\begin{equation}\label{defrho}
\rho(d_B)\,:=\frac{(q-q^{-1})^{l-1}}{l^2}\,\,\sum^{d_B}_{j=1}\sum^l_{s=j+1}\,\,[l-1-j]!
\frac {q^{(j-d_B)(j+1-2s)}} {[j]}\,\times\qquad
\end{equation}
$$
\hphantom{xxxxxxxxxxxxxxxxxxxxxxxx}
\prod_{i=1}^j\!(q-q^{-1})[s-i]^2\,\prod_{i=1}^{j-1}\!\Bigl(q^{(d_B-i)}-
q^{-(d_B-i)}\Bigr)
$$

The main result of the previous calculation - the $SL(2,\bf Z)$ representation
on $\cal V$ - is described in the
next theorem:

\begin{theorem}\label{REP}
Let $P_0,\,N^+_0,\,N^-_0,\,P_1,\dots,N_{(m-1)},\,P_m $  be the ordered  basis
of $\cal V$ as
defined in Lemma \ref{center} . Then the following matrices define a projective
$SL(2,\bf Z)$
representation.

\noindent
The ${\cal T}_o$ matrix is given by (\ref{Tmat}).

\noindent
The ${\cal S}_o$ matrix is given by the following formulae:
\begin{enumerate}
\item For $\,k=0,\ldots,m-1$
$$
{\cal S}_o(N^{\mp}_k)\;\;=\,\;\frac q{\sqrt
l}\,\frac{q-q^{-1}}l[d^{\pm}_k]^2\,\sum_{p=0}^{m-1}\,\frac{[d^{\pm}_kd_p^*]}{[d^*_p]}P_p
\hphantom{xxxxxxxxxxxxxxxxxxxxxxx}
$$
$$
\qquad\qquad\qquad +\; \frac q{\sqrt l}[d^{\pm}_k]^2 d^{\pm}_k
P_m\;\,+\;\,\frac q{\sqrt l}\sum_{p=0}^{m-1}\sum_{\epsilon = \pm
}\,\eta(d^{\pm}_k,d^{\epsilon}_p)N^{\epsilon}_p
\hphantom{xxxxxxxxxxxxxxxxxx}
$$
\item For $\,k=0,\ldots,m-1$
$$
{\cal S}_o(P_k)\,\;=\,\;\frac q{\sqrt l}
\frac{[2d^*_k]}{[d^*_k]}\,P_m\hphantom{xxxxxxxxxxxxxxxxxxxxxxxxxxxxxxxxxxxx}
$$
$$
\qquad\qquad+\;\frac q{\sqrt l}\,\sum_{p=0}^{m-1}\sum_{\epsilon =\pm}\,
\biggl(\mu(d^*_k,d^{\epsilon}_p)+\frac{[2d^*_k]}{[d^*_k]^3}\frac{\eta(d^+_k,d^{\epsilon}_p)+\eta(d^-_k,d^{\epsilon}_p)}{(q-q^{-1})}\biggr)\,N^{\epsilon}_p\;\;
$$
\item
$$
{\cal S}_o(P_m)\,\;=\,\;\frac q{\sqrt l}\,P_m\;+\;\frac q{\sqrt
l}\sum_{p=0}^{m-1}\sum_{\epsilon=\pm}\rho(d^{\epsilon}_p)N^{\epsilon}_p
\hphantom{xxxxxxxxxxxxxxxxxxxxxxx}
$$
\end{enumerate}
 These matrices satisfy the relations
$$
\Bigl({\cal S}_o{\cal T}_o\Bigr)^3\,=\,{\gamma_q}^{-1}q^{(1-m)}\id
$$
$$
({\cal S}_o)^2\,=\,q^2\id
$$
\end{theorem}

Here the superscript $*$ means that either $+$ or $-$ can be inserted yielding
the same result.

\bigskip
\underline{5.) The Structure of the $SL(2,{\bf Z})\,$-Representation on $\cal
V$ :}
\noindent
For small values of $l$ the following polynomial identities hold true:
\beq\label{conjid}
\eta\bigl(d^+_k,d^+_p\bigr)\,+\,\eta\bigl(d^-_k,d^+_p)\,=\,\eta(d^+_k,d^-_p)\,+\,\eta(d^-_k,d^-_p)\quad {\rm and}\quad \quad \rho(\rho^+_p)\,=\,\rho(d^-_p)
\eeq
In this case its easy to see that the representation contains an
$m+1$-dimensional subrepresentation spanned by the $N_j$'s and $P_m\,$.

On this subspace the $\cal T$-matrix
is diagonal and has eigenvalues $ \{\, q^{2j(j+1)},\, j=0,\ldots ,m\}\, $,
i.e., one more than the finite $m$-dimensional representation obtained from the
semisimplified representation category. For prime $l$ it is not hard to see
that the representation is
irreducible. Also, for small $l$ we find that it is finite.

\medskip
It is clear by inspection of th $\cal T$-matrix that a complement to this
representation has to contain the linearly independent nilpotents
$\widetilde{N_j}\,=\,\sum_{\epsilon=\pm}
\frac{d_j^{\epsilon}}{[d_j^{\epsilon}]}N_j^{\epsilon}\;$. Thus it also contains
the
vectors $\,{\cal S}_o\bigl(\widetilde{N_j}\bigr)_k\,$ where the subindex
$k=0,\ldots,m-1\,$ means that we take the component in the $k$-th eigenspace of
$\cal T\,$. Since the elements $\,{\cal S}_o\bigl(\widetilde{N_0}\bigr)_k\,$
are linearly independent from the nilpotents, a $2m$-dimensional complement
exists only if
\beq\label{lindep}
\,{\cal S}_o\bigl(\widetilde{N_j}\bigr)_k\,=\,
c_{jk}{\cal S}_o\bigl(\widetilde{N_0}\bigr)_k\,+\,b_{jk}\widetilde{N_k}\;,
\eeq
for some coefficients $c_{jk}$ and $b_{jk}\,$. Comparison of the coefficients
of the idempotents shows that we need $c_{jk}\,=\,\frac {[d_j]}{[d_k]}
[d_jd_k]\,$, i.e., the
$m\times m$ -matrix $c$ defined by these coefficients is equivalent to the
$\cal S$-matrix
of the semisimple TQFT. We can write polynomial identities similar to
(\ref{conjid})
which are equivalent to (\ref{lindep}) with $b_{jk}=0\,$. Again, for small
values of
$l$ we know that they hold true. Using that ${\cal S}^2$ is proportional to the
identity
they also imply that $\cal S$ decomposes into a tensor product $b\otimes c\,$,
where
$b$ is a two by two matrix with vanishing diagonal elements.
The $\cal T$-matrix on the second summand has eigenvalues
$\{q^{2j(j+1)}\,,\,j=0,\dots,(m-1)\}\,$ all of which are doubly degenerate,
with non trivial Jordan-bloc. For a suitable
normalization we thus expect the second summand to be the tensorproduct of the
two dimensional standard representation and the known $m$-dimensional finite
representation.

\medskip
In a TQFT  $\widetilde{SL(2, {\bf Z})}\,$ extends to
representations of modular groups at higher genus. If these factor through
their actions
on the homology of the surface the projective $SL(2,{\bf Z})\,$-representation
extends  to representations of higher symplectic group. It is a fact that for
congruence groups as the higher symplectic groups over $\bf Z$ any irreducible
representation is the tensorproduct of a finite and an algebraic
representation, see [Kz]. Thus it is
likely that the tensorproduct presentation described in the previous paragraph
can
also be inferred from rather general arguments.

\medskip
 We summarize our observations in the following conjecture. In the next section
we show that it holds true for the five and seven dimensional representation.

\begin{guess}
The projective, $3m+1$ - dimensional $SL(2,\bf Z)$ representation
defined in Theorem (\ref{REP}) decomposes as
$$
{\cal V}\,=\,{\cal V}_N\,\oplus\,{\cal V}_{stan}\otimes {\cal V}_{semis}
$$
where
\begin{enumerate}
\item ${\cal V}_N$ is an $(m+1)$ - dimensional, irreducible, finite
representation spanned by $\,N_j\,=\,N_j^++N_j^-\,$ and
$P_m\,$, see e.g. (\ref{N2}) or (\ref{7N3}).
\item ${\cal V}_S$ is the $2m$ - dimensional subrepresentation spanned by
$$\widetilde{N_j}\,=\,\sum_{\epsilon=\pm}\frac{d_j^{\epsilon}}{[d_j^{\epsilon}]}N_j^{\epsilon}\;
$$
and the $j$-th ${\cal T}$ - eigenspace components
$$
\bigl({\cal S}_o(\widetilde{N_0})\bigr)_j\;.
$$
This representation is the tensorproduct
$
{\cal V}_S\,=\,{\cal V}_{stan}\otimes {\cal V}_{semis}
$
of
\ben
\item the two dimensional, algebraic standard representation ${\cal V}_{stan}$
as in (\ref{S2}) or (\ref{7stan}) and
\item  an $m$ -dimensional finite representation   ${\cal V}_{semis}$
 which is isomorphic - up to a projective phase - to the $SL(2,\bf Z)$
representation obtained from the semisimple subquotient category, see for
example (\ref{7semis}).
\een
\end{enumerate}
\end{guess}

\noindent
\underline{6.) The Examples $l\,=\,3,\,5\;$} :
In this section we verify the conjecture of the previous section for $l=3$ and
$l=5\,$.
We compute the explicit representation matrices of the various finite
representations:

\medskip
For $l=3$ the matrices of the  $SL(2,\bf Z)$ are given in the
basis$P_0,\,N_0^+,\,N^-_0,\,P_1\,$ by:

\begin{equation}\label{TN3}
{\cal T}_o\,=\,\quad\left[\begin{array}{cccc}
  1  &\qquad  0    &\qquad  0&\qquad  0\\
&&&\\
1  &\qquad  1    &\qquad  0&\qquad  0\\
&&&\\
-2  &\qquad  0    &\qquad  1 &\qquad  0\\
&&&\\
 0    &\qquad  0&\qquad   0  &\qquad  q
\end{array}
\right]
\hphantom{xxxxxxxxxxxxxxxxxx}
\end{equation}

\bigskip

\begin{equation}\label{S3}
{\cal S}_o\,=\,\frac q {\sqrt 3}\left[\begin{array}{cccc}
0  & -\frac 1 3 (q-q^{-1}) &\frac 1 3 (q-q^{-1})&0 \\
&&&\\
\frac 2 3 (q-q^{-1})& -1 & 0 &-\frac 2 3 (q-q^{-1})\\
&&&\\
-\frac 7 3 (q-q^{-1})& -1 & 0 &-\frac 2 3 (q-q^{-1})\\
&&&\\
-1 & \frac 2 3 (q-q^{-1})&\frac 1 3 (q-q^{-1})&1

\end{array}
\right]
\end{equation}

This representation decomposes into the sum of two irreducible,
two-dimensional subrepresentations
$$
{\cal V}\;=\;{\cal V}_N\,+\,{\cal V}_S\;.
$$
Here the subspace ${\cal V}_N$ is spanned by $N_0\,=\,N_0^+\,+N_0^-$ and $P_1$
with $\cal S$ and ${\cal T}$ acting as:

\begin{equation}\label{N2}
{\cal T}_N\,=\,\left[\begin{array}{cc}
  1  &\qquad  0   \\
&\\
  0  &\qquad  q
\end{array}
\right]\qquad{\cal S}_N\,=\,\frac q {\sqrt 3}
 \left[\begin{array}{cc}
  -1  &  -\frac 2 3 (q-q^{-1})   \\
&\\
 ( q-q^{-1}) &  1
\end{array}
\right]
\end{equation}
This subrepresentation ${\cal V}_S$ has basis vectors $\widetilde{\, P_o}\,=\,
P_o\,+\,\frac 1 {(q-q^{-1})}N_o\,$ and $\widetilde {\,N_o}\,:=\,
N^+_o-2N^-_o\;$ for which the $\cal S$ and ${\cal T}$ matrix have the form of
the standard representation:

\begin{equation}\label{S2}
{\cal T}_S\,=\,\left[\begin{array}{cc}
  1  &\qquad  0   \\
&\\
  1  &\qquad  1
\end{array}
\right]\qquad{\cal S}_S\,=\,q \gamma_q \left[\begin{array}{cc}
  0  &\qquad -1  \\
&\\
1 & \qquad 0
\end{array}
\right]
\end{equation}

\bigskip

 For $l=5$ the matrices are given for the basis
$P_0,\,N^+_0,\,N^-_0,\;P_1\,N^+_1,\,N^-_1,\;P_2\,$:
\begin{equation}\label{T5}
\hspace{-2.5cm}{\cal T}_o\,=\,\left[\begin{array}{ccccccccc}
1&\qquad 0&\qquad 0&\vdots&\qquad 0&\qquad 0&\qquad 0&\vdots&\qquad
0\\&&&\vdots&&&&\vdots&\\
1&\qquad 1&\qquad 0&\vdots&\qquad 0&\qquad 0&\qquad 0&\vdots&\qquad
0\\&&&\vdots&&&&\vdots&\\

-4&\qquad 0&\qquad 1&\vdots&\qquad 0&\qquad 0&\qquad 0&\vdots&\qquad
0\\\cdots\,\,\cdots&\cdots\,\,\cdots&\cdots\,\,\cdots&\vdots&\cdots\,\,\cdots&\cdots\,\,\cdots&\cdots\,\,\cdots&\vdots&\cdots\,\,\cdots\\
0&\qquad 0&\qquad 0&\vdots&\qquad q^{-1}&\qquad 0&\qquad 0&\vdots&\qquad
0\\&&&\vdots&&&&\vdots&\\

0&\qquad 0&\qquad 0&\vdots&\qquad  -q^{-1}\frac 3 {[2]}  &\qquad q^{-1}
&\qquad 0&\vdots&\qquad 0\\&&&\vdots&&&&\vdots&\\
0&\qquad 0&\qquad 0&\vdots&\qquad q^{-1}\frac 2 {[2]} &\qquad 0&\qquad
q^{-1}&\vdots&\qquad
0\\\cdots\,\,\cdots&\cdots\,\,\cdots&\cdots\,\,\cdots&\vdots&\cdots\,\,\cdots&\cdots\,\,\cdots&\cdots\,\,\cdots&\vdots&\cdots\,\,\cdots\\
0&\qquad 0&\qquad 0&\vdots&\qquad 0&\qquad 0&\qquad 0&\vdots&\qquad q^2

\end{array}
\right]
\end{equation}

\begin{equation}\label{S5}
\hspace{-3.2cm}{\cal S}_o\,=\,\frac q {\sqrt 5}
\,\left[\begin{array}{ccccccccc}

0& - \xi &   \xi
                  &\vdots &  0     &\xi(-1+2[2])&\xi(1-2[2])
                                                &\vdots & 0
\\&&&\vdots&&&&\vdots&\\

\xi (2+4[2]) &  [2]&  0
                  &\vdots &  \xi(2-6[2])  & \frac 1 5 (-2+[2]) &\frac 1 5
(2-6[2])

                                                 &\vdots & \xi (-4+2[2])
\\&&&\vdots&&&&\vdots&\\

\xi (4[2]-23) & [2]& 0
             &\vdots &\xi (-23+19[2])&  \frac 1 5 (-2+[2])     &\frac 1 5
(2-6[2])
                                                 &\vdots &\xi (-4+2[2])\\
\cdots\,\,\cdots&\cdots\,\,\cdots&\cdots\,\,\cdots&\vdots&\cdots\,\,\cdots&\cdots\,\,\cdots&\cdots\,\,\cdots&\vdots&\cdots\,\,\cdots\\

0&-\xi&   \xi
             &\vdots & 0 & -\xi[2]&\xi[2]
                                                  &\vdots & 0
\\&&&\vdots&&&&\vdots&\\

-\xi (12+14[2])  & -3-2[2]    & 0
           &\vdots & \xi(18+16[2])& \frac 1 5 (-2+[2])  & \frac 1 5(2+4[2])
                                                     &\vdots &-\xi(6+2[2])
\\&&&\vdots&&&&\vdots&\\

 \xi (13+11[2])   & -3-2[2]    & 0
              &\vdots & -\xi(7+9[2])& \frac 1 5 (-2+[2]) &  \frac 1 5(2+4[2])

                                                        &\vdots &
-\xi(6+2[2])\\

\cdots\,\,\cdots&\cdots\,\,\cdots&\cdots\,\,\cdots&\vdots&\cdots\,\,\cdots&\cdots\,\,\cdots&\cdots\,\,\cdots&\vdots&\cdots\,\,\cdots\\

[2] &  4\xi  &\xi &\vdots &
                     -1-[2]& \xi(2-2[2]) & \xi(3-3[2])
                                                       &\vdots & 1

\end{array}
\right]
\end{equation}
\medskip

 where
$$
\qquad\qquad \xi\,=\,\frac {q-q^{-1}} 5 \quad .
$$

This representation decomposes into two irreducible representations ${\cal
V}_N$
and ${\cal V}_S$, with $dim({\cal V}_N  )=3$ and  $dim({\cal V}_S  )=4$. The
three dimensional representation is spanned by the vectors

$$ N_0\,=\,\frac 1 {q-q^{-1}}( N_0^+\,+\,N_0^-)\quad
N_1\,=\,  \frac 1 {q-q^{-1}}( N_1^+\,+\,N_1^- )\quad\;{\rm and}\quad P_2\;.
$$

 The representation matrices are

\begin{equation}\label{7N3}
\hspace{-2.2cm}
{\cal T}_N\,=\,\left[\begin{array}{ccc}
  1  &\qquad  0    &\qquad  0\\
&&\\
0 &\qquad  q^{-1}    &\qquad  0\\
&&\\
0  &\qquad  0    &\qquad  q^2
\end{array}
\right]
\qquad
{\cal S}_N\,=\,\frac q {\sqrt 5}
\left[\begin{array}{ccc}
[2]  & -[2] & 2 \\
&&\\
-(3+2[2])& [2] & (4+2[2]) \\
&&\\
1&  (1-[2])&1
\end{array}
\right]
\end{equation}

The four dimensional representation is spanned by
$$
\widetilde {P_0}\,:=\,P_0 -[2]N_0\qquad\widetilde {P_1}\,:=\,\frac
1{[2]}\,\bigl(P_1 +(3+2[2])N_1\bigr)
$$
and
$$
\widetilde {N_0}\,:=\,(N^+_0\,-4N^-_0 )\qquad \widetilde {N_1}\,:=\frac
1{[2]^2}(-3N^+_1\,+2N^-_1)\;.
$$
With ordering $\widetilde {P_0},\,\widetilde {N_0},\,\widetilde {P_1},\,
\widetilde {N_1}$ we find the matrices:

\begin{equation}\label{7S4}
\hspace{-1.8cm}
{\cal T}_S\,=\,\left[\begin{array}{ccccc}
  1  &  0 &\vdots&0   &  0\\
& &\vdots& &\\
1 &1 &\vdots&0&0\\
\cdots&\cdots &\vdots&\cdots&\cdots \\
0  &  0  &\vdots& q^{-1}  & 0 \\
& &\vdots& &\\
0  &  0  &\vdots& q^{-1}  & q^{-1} \\
\end{array}
\right]
\qquad
{\cal S}_S\,=\,\frac {q(q-q^{-1})} {\sqrt 5}
\left[\begin{array}{ccccc}
0&-1&\vdots&0   &-[2] \\
& &\vdots&  &\\
1&0&\vdots&[2] &0\\
\cdots&\cdots &\vdots&\cdots&\cdots \\
0&-[2]&\vdots&0   &1 \\
& &\vdots&   & \\
{[2]}  & 0 & \vdots & -1 &0\\

\end{array}
\right]
\end{equation}

Now it is easy to see that this can be written as a tensorproduct of
$SL(2,\bf Z)$ representations:
$$
{\cal V}_S\,\cong\,{\cal V}_{stan}\otimes{\cal V}_{semis}\;,
$$
In order to denote the isomorphism
$$
\widetilde {P_i}\,\to \,v_P \otimes w_i\qquad\qquad\qquad \widetilde
{N_i}\,\to\, v_N \otimes w_i
$$
we introduce bases $\{v_P,v_N\}$ and $\{w_0,w_1\}$ of  ${\cal V}_{stan}$ and
${\cal V}_{semis}$ respectively. For these bases we can write

$$
{\cal T}_N\,=\,{\cal T}_{stan}\otimes {\cal T}_{semis}\qquad\qquad\qquad
{\cal S}_N\,=\,q{\cal S}_{stan}\otimes{\cal S}_{semis}\qquad
$$

with

\begin{equation}\label{7stan}
{\cal T}_{stan}\,=\,\left[\begin{array}{cc}
  1  &\qquad  0   \\
&\\
  1  &\qquad  1
\end{array}
\right]
\qquad
\qquad
{\cal S}_{stan}\,=\,\left[\begin{array}{cc}
  0  &\qquad -1  \\
&\\
1 & \qquad 0
\end{array}
\right]
\end{equation}

and

\begin{equation}\label{7semis}
{\cal T}_{semis}\,=\,\left[\begin{array}{cc}
  1  &\qquad  0   \\
&\\
  0  &\qquad  q^{-1}
\end{array}
\right]
\qquad{\cal S}_{semis}\,=\,\frac {(q-q^{-1})} {\sqrt 5}\left[\begin{array}{cc}
  1  &\qquad [2]  \\
&\\
{[2]} & \qquad -1
\end{array}
\right]
\end{equation}

\section*{Conclusion}
The results of Chapter 2 show that a the construction of a universal TQFT
should include two features. One is to avoid degeneracies by considering
only doubles. The fact that the projective phases and the proofs of modular
relations are
most conveniently given in terms the bilinear forms and moduli defined from the
integrals is an indication that this is the correct language also for
constructions at higher genus. In view of the glueing operations described in
the introduction the genus one case can in fact be thought of as a basic
building bloc.
It should be possible to understand more conceptually the appearance of the
finite representation we know from
the semisimple theory as a tensorproduct with the standard representation
rather
than a sub representation. In particular it should be interesting to see
how the representation on general $\dA\,$ is modified if we pass to the
semisimple quotient of the representation category of $\dA\,$ and a possible
truncation of the resulting TQFT.

Also, the appearance of algebraic representations is a novel feature of these
theories. We expect to find higher dimensional algebraic representations of
$SL(2,Z)\,$ if we start from higher rank quantum groups for which the orders of
nilpotencies of central elements will be higher.

\bigskip
\bigskip

\section*{References}

[A] Atiyah,M.:``Topological Quantum Field Theories", Publ. Math. IHES {\bf 68}
(1989) 175-186.

[Ab] Abe, E.:``Hopf Algebras", Cambridge University Press (1977).

[Cr] Crane,L.:``2-d Physics and 3-d Topology",  Commun. Math. Phys. {\bf 135}
(1991) 615-640.

[D] Deligne, P.:``Cat\'egories Tannakiennes", The Grothendieck Festschrift,
Vol. II, Progress in Mathematics, Birkh\"auser, Boston (1990).

[Dr0] Drinfel'd,V.G.:``Quantum Groups", Proc. Internat. Congr. Math., Berkeley,
1986, Am. Math. Soc. , Providence, R.I. (1987) 798-820.

[Dr1] Drinfel'd, V.G.:``On Almost Cocommutative Hopf Algebras", Leningrad Math.
J., {\bf 1} (1990)No.2, 321-342 .

[K] Kerler,T.:``Non-Tannakian Categories in Quantum Field Theory", New Symmetry
Principles in Quantum Field Theory. NATO ASI Series B: Physics  Vol. {\bf 295},
Plenum Press (1992).

[KR] Kazhdan, D., Reshetikhin,N.Y.: privat communication. Harvard lecture
notes.

[Kz] Kazhdan, D.:``On Arithmetic Varieties", in ``Lie Groups and Their
Representations", Adam Hilger LTD, London, (1971).

[L] Lang,S.:``Algebraic Number Theory", Addison-Wesley, New York (1970).

[LSw] Larson, R.G., Sweedler, M.E.:``An Associative Orthogonal Bilinear Form
for Hopf Algebras", Amer. J. of Math., {\bf 91},(1969) No 1., 75-94.

[Ly] Lyubashenko,V.:``Tangles and Hopf Algebras in Braided Tensor Categories",
to appear in J. Pure Appl. Alg.

Lyubashenko,V.:``Modular Transformations for Tensor Categories", to appear on
J. Pure Appl. Alg.

[LyM] Lyubashenko, V.,Majid,S.:``Braided Groups and Quantum Fourier Transform",
(1991) to appear in J.Algebra.

[M1] Majid, S.:``Braided Matrix Structure of the Sklyanin Algebra and of the
Quantum Lorentz Group", Commun. Math. Phys. {\bf156} (1993) 607-638.

[M2] Majid, S.:``Braided Groups", J. Pure Appl. Alg.  {\bf86} (1993) 187-221.

[Mc] MacLane,S.:``Categories for the Working Mathematician",Springer-Verlag,
1971.

[Rd] Radford,D.E.:``The Antipode of a Finite Dimensional Hopf Algebra is
Finite", Amer. J. Math. {\bf 98} (1976) 333-355.

[RT0] Reshetikhin,N.Y., Turaev,V.:``Ribbon Graphs and Their Invariants Derived
{}From Quantum Groups", Commun. Math. Phys. {\bf 127} (1990) 1-26.

[RT] Reshetikhin, N.Y., Turaev, V.:``Invariants of 3-Manifolds via Link
Polynomials and Quantum Groups", Invent. Math {\bf 103} (1991) 547-598.

[T] Turaev, V.:``Quantum Invariants of 3-Manifolds",  Publication de l'Institut
de Recherche Math\'ematique Avanc\'ee, Strasbourg, Preprint: ISSN 0755-3390.

[TV] Turaev,V., Viro, O.:``State Sum Invariants of 3-Manifolds and Quantum
6-j-Symbols'', Topology {\bf 31} (1992) 865-902.

[Wi]Witten,E.:``Quantum Field Theory and the Jones Polynomial", Commun. Math.
Phys.
{\bf 121} (1989) 351-399.

\end{document}